\newif\ifShowKeys
\ShowKeysfalse

\documentclass[11pt,a4paper]{article} 
\pdfoutput=1

\usepackage{booktabs} 
\usepackage{colortbl} 
\usepackage{xcolor}

\usepackage[no-natbib-sort]{my-jheppub}
\usepackage{amsmath, amssymb}

\usepackage{mathpazo}
\usepackage{bm}
\usepackage{environ}
\usepackage{mathrsfs}
\usepackage{array,arydshln}

\usepackage{graphicx,epsfig}
\usepackage{tikz}							% necessary for conformal blocks
\usepackage{epic}
\usepackage{youngtab}
\usepackage{float}
\usepackage{color}
\definecolor{maroon}{rgb}{0.8,0.3,0.}

\usepackage{slashed}
\usepackage[nodayofweek]{date time}

\ifShowKeys \usepackage[notcite]{showkeys} \fi

\usepackage{hyperref}

\usepackage{aurical}
\usepackage[T1]{fontenc}

\usepackage{framed}
\definecolor{shadecolor}{RGB}{255, 230, 204}

\allowdisplaybreaks

\newcommand{\be}{\begin{equation}}
\newcommand{\ee}{\end{equation}}

\newcommand{\mc}{\mathcal }

\newcommand{\bv}{\begin{verbatim}}
\newcommand{\ev}{\end{verbatim}}

\newcommand{\la}{\label}

\newcommand{\eps}{\varepsilon}

\newcommand{\E}{\text{E}}

\def\confblock(#1,#2,#3,#4,#5,#6,#7) { #1_{#2}\!
{\small\begin{bmatrix} #4 & #5 \\ #3 & #6  \end{bmatrix}}\!(#7) }

{}

\newcommand{\vev}[1]{\left\langle #1 \right\rangle}
\newcommand{\tr}{\text{Tr}\,}
\newcommand{\T}[1]{\langle \text{Tr}\, \varphi^{#1} \rangle}
\newcommand{\Chbar}{\mathcal{C}_{\hbar}}
\newcommand{\p}{\text{p}}
\newcommand{\s}{\text{s}}

\title{Chiral trace relations in $\Omega$-deformed $\mc N=2$ theories}
\author[a,b]{Matteo Beccaria,} 
\author[a,b]{Alberto Fachechi,} 
\author[a,b]{Guido Macorini} 

\abstract{
We consider $\mc N=2$ $SU(2)$ gauge theories in four dimensions (pure or mass deformed)
and discuss the properties of the simplest chiral observables in the presence of a generic 
$\Omega$-deformation. 
We compute them by equivariant localization and 
analyze the structure of the exact instanton corrections to the classical 
chiral ring relations. We predict exact relations valid at all instanton number
among the traces $\langle\text{Tr}\varphi^{n}\rangle$, where $\varphi$ is the scalar field in the gauge
multiplet. In the Nekrasov-Shatashvili limit, such relations may be explained in terms of the 
available quantized Seiberg-Witten curves. 
Instead, the full two-parameter deformation enjoys novel 
features and the ring relations require non trivial additional derivative terms 
with respect to the modular parameter. Higher rank groups are  briefly discussed emphasizing 
non-factorization of correlators due to the $\Omega$-deformation. 
Finally, the structure of the deformed ring relations
in the $\mc N=2^{\star}$ theory is analyzed from the point of view of the Alday-Gaiotto-Tachikawa
correspondence proving  consistency as well as some interesting universality properties.
\vfill }

\affiliation[a]{Dipartimento di Matematica e Fisica Ennio De Giorgi,\\
Universit\`a del Salento, Via Arnesano, 73100 Lecce, 
Italy} 

\affiliation[b]{INFN, Via Arnesano, 73100 Lecce, Italy}

\emailAdd{matteo.beccaria@le.infn.it} 
\emailAdd{alberto.fachechi@gmail.com} 
\emailAdd{macorini@nbi.ku.dk} 

\begin{document}

%\date{\currenttime}
%\begin{flushleft}\boxed{\small{\tt \today \ \ - \ \  \currenttime }}\end{flushleft}

% \begin{flushright}\small{Imperial-TP-AT-2015-{06}}\end{flushright}				% report number

\maketitle
\flushbottom

\section{Introduction and results}

Four dimensional $\mathcal{N} = 2$ super Yang-Mills (SYM) theories 
are a unique theoretical laboratory where non-perturbative effects are fully under
control. This is achieved by combining the Seiberg-Witten (SW) description of the low-energy 
effective theory \cite{Seiberg:1994rs,Seiberg:1994aj} with the localization 
computation of instanton corrections 
\cite{Nekrasov:2002qd,Flume:2002az,Nekrasov:2003rj,Bruzzo:2002xf, Losev:2003py,Flume:2004rp}. 
Additional structure is available in superconformal theories with possible 
mass deformations \cite{Teschner:2014oja}. In this case, a large variety of  new tools 
have been developed as, in particular, the relation to integrable models \cite{Nekrasov:2009rc}
and the Alday-Gaiotto-Tachikawa (AGT) 2d/4d  correspondence \cite{Alday:2009aq,Alday:2009fs}.
%matrix model techniques \cite{Dijkgraaf:2009pc,Cheng:2010yw}, and the relation with 
%topological string amplitudes through geometric engineering 
%\cite{Antoniadis:2010iq,Huang:2011qx,Florakis:2015ied}. 

\medskip
The duality properties of these models determine important constraints \cite{Gaiotto:2009we}.
Remarkable results may be obtained in $\mathcal{N}=2^\star$ theories
where an adjoint hypermultiplet of mass $m$ is present beside the gauge vector multiplet. 
The mass deformation interpolates the $\mathcal{N} = 4\to \mathcal{N} = 2$ transition.
The $S$-duality of the $\mathcal{N} = 4$ theory is inherited for generic mass
and the prepotential is constrained by a modular anomaly equation 
\cite{Minahan:1997if}. \footnote{
The modular anomaly constraint  appears in several contexts, {\em e.g.} 
in an $\Omega$ background
\cite{Huang:2006si,Grimm:2007tm,Huang:2009md,Huang:2010kf,Huang:2011qx,
Galakhov:2012gw,Billo:2013fi,Billo:2013jba,Nemkov:2013qma,Billo:2014bja, 
Lambert:2014fma, Beccaria:2016wop,Beccaria:2016nnb,Beccaria:2016vxq}, 
 from the point of view of the AGT correspondence
\cite{Marshakov:2009kj,KashaniPoor:2012wb,Kashani-Poor:2013oza,Kashani-Poor:2014mua}, 
in the large-$N$ limit \cite{Billo:2014bja}, and in SQCD models with fundamental matter 
\cite{Billo:2013fi,Billo:2013jba,Ashok:2015cba,Ashok:2016oyh}. 
Recently, it has been shown to be associated with the all-order WKB expansion 
of quantum mechanical models \cite{Codesido:2016dld,Basar:2017hpr}.
}
As a consequence of this structure, it is possible to encode the mass expansion of the 
prepotential in terms of quasi-modular functions of the gauge coupling and the vacuum expectation 
$\vev{\varphi}$ of the scalar in the gauge multiplet. The construction is general and holds
for arbitrary gauge groups \cite{Billo:2015pjb,Billo:2015jyt,Billo:2016zbf}.

\medskip
Recently, the authors of \cite{Ashok:2016ewb} have presented a  detailed study of the 
modular properties of  specific observables different from the prepotential, {\em i.e.}
the chiral traces $\T{n}$ in ${\mathcal N} =2^\star$ $U(N)$ gauge theories, where $\varphi$
is the scalar field in the gauge multiplet. 
Supersymmetry implies that correlators of chiral operators factorize and can 
be expressed in terms of such traces that parametrize the (quantum) chiral ring.
%see also \cite{Gerchkovitz:2016gxx} for more general cases. 
An important  motivation for the analysis of these chiral observables
is that they play a  role in the physics of surface operators
 \cite{Gukov:2014gja,Gukov:2006jk,Gukov:2008sn}. The associated infrared effects 
 are indeed captured  by a twisted two dimensional effective superpotential
 that  can be computed from the expectation values of (higher order) 
 chiral ring elements in the bulk four dimensional theory \cite{Gaiotto:2013sma}.
 
 \medskip
In our analysis, we present a discussion of special relations between higher order 
traces that are exact at all orders in the instanton expansion. In 
the simplest case of the $SU(2)$ gauge group,  these are expressions
 for $\T{n}$ in terms of $\T{2}$ for generic parameters of the gauge theory, {\em i.e.}
 the hypermultiplet mass $m$ and the components of $\vev{\varphi}$.
 At the classical level, higher order traces are not independent quantities due to elementary properties
of $SU(2)$ matrices. Thus, for even $n$ (odd traces are zero)
\be
\la{1.1}
\T{2n} = P_{n}({u}), \qquad {u}=\T{2},
\ee
where $P_{n}$ is a degree $n$ polynomial in ${u}$ with constant coefficients. \footnote{
Eq. (\ref{1.4})  is clearly nothing but an elementary relation among symmetric polynomials.}
In the following, we shall call
{\em trace~relations} constraints like (\ref{1.1}). Non perturbative instanton effects 
introduce non-trivial corrections. In the pure gauge theory, the
quantum trace relations read 
\be
\la{1.2}
\T{2n} = P_{n}({u}, q),
\ee
where now $P_{n}$ is a new polynomial in ${u}$ with coefficients depending polynomially
on the instanton counting parameter $q$ and computable from the gauge theory  resolvent
\cite{Cachazo:2002ry, Flume:2004rp, Nekrasov:2012xe}. 
Moving to the $\mc N=2^{\star}$ theory, the structure of instanton corrections is completely
different. It is again possible to write 
\be
\la{1.3}
\T{2n} = P_{n}^{\star}({u}, q),
\ee
but now $P_{n}^{\star}$ is a polynomial in ${u}$ whose coefficients may be expressed as
polynomials in certain Eisenstein series $\E_{n}(q)$ \cite{Fucito:2005wc}. 
Thus, there are still exact 
trace relations, but the precise instanton dependence is more involved as predicted by $S$-duality
\cite{Ashok:2016ewb}. Extension to higher rank groups is essentially the same, except for
a larger number of independent coordinates playing the role of ${u}$.

\medskip
It is quite interesting and natural to investigate the properties of chiral traces in 
$\Omega$-deformed $\mc N=2$ theories. This 4d Poincar\' e breaking deformation 
depends on two parameters $\eps_{1}, \eps_{2}$ and is the main ingredient in the 
 localization approach of 
\cite{Nekrasov:2002qd,Nekrasov:2003rj,Nakajima:2003uh} where it 
is needed to regularize
the multi-instanton moduli space 
\cite{Flume:2002az,Bruzzo:2002xf,Flume:2004rp,Nekrasov:2004vw,Marino:2004cn,
Billo:2009di,Fucito:2009rs,Billo:2010bd}. \footnote{
The  string interpretation of the $\Omega$-background and its BPS excitations
has been fully clarified in 
\cite{Hellerman:2011mv,Hellerman:2012zf,Hellerman:2012rd,Orlando:2013yea} where 
a geometric interpretation of  localization 
in terms of a dilaton potential is provided giving also  a clear picture 
of all possible deformation parameters.
}
The $\Omega$-deformed  prepotential expanded around the 
undeformed limit is the generating function of higher genus amplitudes of the $\mc N=2$
topological string  \cite{Antoniadis:1993ze,Antoniadis:2010iq,Krefl:2010fm,
Huang:2010kf,Antoniadis:2013mna,Antoniadis:2013epe,Florakis:2015ied} and 
satisfies a 
holomorphic anomaly 
equation \cite{Bershadsky:1993ta,Bershadsky:1993cx,Klemm:2002pa,Huang:2009md}.
At finite  $\eps_{1}, \eps_{2}$, the (deformed) partition function is also a 
fundamental object  within the AGT correspondence 
\cite{Alday:2009aq}. In this context, the deformed $\mc N=2$  instanton partition function
is mapped to  conformal blocks of a suitable CFT  and  may be tested perturbatively in the 
instanton number \cite{Poghossian:2009mk} and proved in various 
cases like $\mc N=2^{\star}$ theory \cite{Fateev:2009aw} or linear quivers on sphere
\cite{Alba:2010qc}. \footnote{
In the deformed  
$\mc N=2^{*}$  $SU(2)$ gauge theory the CFT dual quantity is the one-point 
conformal block on the torus. In the $SU(2)$ theory with four fundamental
flavours, it is the conformal block of four Liouville operators on a sphere
\cite{Fateev:2009me,Hadasz:2009db,Menotti:2010en,Menotti:2011ws,Menotti:2012wq,
Menotti:2014kra,Menotti:2016prf,Menotti:2016jut,
Marshakov:2010fx,
KashaniPoor:2012wb,Piatek:2013ifa,Kashani-Poor:2013oza,
Kashani-Poor:2014mua,Alkalaev:2016ptm}.
The pure gauge case is somewhat special and has been discussed in \cite{Gaiotto:2009ma}.}

\medskip
 In the simplest case of the $SU(2)$ gauge group the analysis of the $\eps$-deformed chiral ring
 aims at finding suitable generalizations of (\ref{1.2}) and (\ref{1.3}) taking into account the 
 deformation parameters $\eps_{1}$, $\eps_{2}$. Notice that a priori it is not at all trivial 
 that such generalization exists in some reasonable simple form. In 
principle, this issue may be addressed by (at least) three different approaches. 
The first is based on
the use of  a quantized SW curve taking into 
account the $\eps$-deformation. Investigations in this direction have been discussed 
in \cite{Fucito:2011pn,Poghossian:2010pn}. Another option is to exploit 
the topological string in the spirit of \cite{Klemm:2002pa,Hollowood:2003cv,Eguchi:2003sj}.
Finally, one can exploit AGT correspondence. Very recently, this approach has been 
applied in  the $N_{f}=4$ theory \cite{Fucito:2015ofa, Fucito:2016jng}. Important
differences may be  expected in the $\mc N=2^{\star}$ theory due to its peculiar modular structure. 
Reversing the logic, an explicit microscopic computation of the relations between the traces
$\T{n}$ may be used to 
prove the correctness of any proposal for the deformed SW curve, or as a test of AGT correspondence. 

\medskip
We  looked for the existence of such relations 
by inspecting the explicit expressions of 
$\T{n}$ computed from localization  at high instanton number in various 
deformed $\mc N=2$ theories.
The results of our analysis show that the undeformed trace relations 
undergo important modifications when the $\Omega$-background is switched on. 
To illustrate the changes in (\ref{1.2}) and (\ref{1.3}) it is convenient to provide some simple
illustrative example. In pure   $\mc N=2$ $SU(2)$ gauge theory, 
a prototypical relation we obtain is 
\be
\la{1.4}
\T{4} = \frac{1}{2}\,{u}^{2}+4\,q-\eps_{1}\,\eps_{2}\,{u}',
\ee
where ${u} = \T{2}$ and $X' = q\,\partial_{q}\,X$ where $q$ is the instanton counting parameter.
The undeformed limit contains the well known one-instanton correction to the classical 
chiral ring. The novel
term, last in (\ref{1.4}), is present for $\eps_{1}\eps_{2}\neq 0$, so it vanishes in the 
Nekrasov-Shatashvili (NS) limit  \cite{Nekrasov:2009rc} 
where one of the two $\eps$ parameters
vanishes. In general, for higher order traces, the gravitational corrections survive even in the NS
limit. An example, still in pure gauge theory, is the 6-th order trace relation
\be
\la{1.5}
\T{6} = \frac{1}{4}\,{u}^{3}+6\,q\,{u}-\frac{3}{2}\,\eps_{1}\,\eps_{2}\,{u}\,{u}'+
\eps_{1}^{2}\, \eps_{2}^{2}\,{u}''+6\,q\,(3\,\eps_{1}^{2}+4\,\eps_{1}\,\eps_{2}+3\,\eps_{2}^{2}),
\ee
where the last term remains if only one of the two $\eps$ parameters vanishes.
In general, the NS limit of results like (\ref{1.5}) may be treated perturbatively in the instanton
number by means of 
the available deformed SW curves. However, the modular structure is not captured in an automatic
way, although modularity is still at work. 
To explain this point, we can consider the simplest NS trace relation in the
$\mc N=2^{\star}$ theory. We shall show that for $(\eps_{1}, \eps_{2})=(\hbar, 0)$
\be
\la{1.6}
\T{4} = \frac{{u}^2}{2}+\frac{\Chbar}{12}\, {u} \,(\,\E_2-1)
-\frac{\Chbar}{1440}\,\bigg[\Chbar\, \left(5 \,\E_2^2-5 \,\E_2-\,\E_4+1\right)+12 (\,\E_4-1)\, 
\hbar ^2\bigg],
\ee
where $\Chbar = 4 m^2-\hbar^{2}$. 
The relation (\ref{1.6}) is exact, {\em i.e.} valid at all instanton numbers. The instanton parameter $q$
is the argument of the Eisenstein series $\E_{k}\equiv \E_{k}(q)$. It seems 
non trivial to extend the undeformed methods of  \cite{Fucito:2005wc} to get relations 
like (\ref{1.6}) in a systematic way.
Even worse, in the fully deformed case, with $\eps_{1}\eps_{2}\neq 0$, there is no obvious 
way to construct a doubly deformed curve providing the $q\partial_{q}$
derivative terms in (\ref{1.4}) and (\ref{1.5}).

\medskip
Remarkably, we show that in the AGT perspective these extra terms
are instead natural. Loosely speaking, they turn out to be
associated with higher powers of the Virasoro operator $L_{0}$ in the Liouville integrals of motion. 
This means that  the  trace relations for the $\mc N=2$ theories 
on a generic $\Omega$ background may be considered as a test of AGT, in the same way as 
they were in the $N_{f}=4$ theory considered in \cite{Fucito:2015ofa, Fucito:2016jng}.
The  identification of the chiral observables $\T{n}$ with the conserved hamiltonians
of the Liouville integrable system is of course expected a priori \cite{Nekrasov:2009rc},
 although its precise 
dictionary requires a choice of basis. In general terms, a simple nice consequence of the 
AGT interpretation is the prediction of definite universality
 properties of certain  leading derivative terms in the trace relations. 

\medskip
The plan of the paper is the following. 
We begin by discussing undeformed trace relations  and their proof 
in Section \ref{sec:undef}. Gravitational corrections, {\em i.e.} the dependence on the 
$\eps$ deformation, are presented in Section \ref{sec:gravpure} in the pure $\mc N=2$
gauge theory. The $\mc N=2^{\star}$ theory is considered in Section \ref{sec:star-NS}
by looking first at the NS limit, and later in Section \ref{sec:star-full} by switching on 
a general background. Section \ref{sec:un} is devoted to the generalization of the previous
discussion to $U(N)$ gauge groups. There are no remarkable differences with the exception of 
interesting violation to supersymmetric factorization of correlators. This is expected in the 
deformed theory \cite{Fucito:2005wc} and we  give explicit examples.
Finally, Section \ref{sec:agt} presents a discussion of the trace relations in the $\mc N=2^{\star}$
theory from the point of view of the AGT correspondence checking the predicted universality 
property of some derivative terms. Several Appendices present 
technical details, tools,    and side comments.

\section{Trace relations in undeformed theories}
\la{sec:undef}

In this section, we consider the undeformed pure gauge $SU(2)$ $\mc N=2$ theory and its 
$\mc N=2^{\star}$ mass deformation. We begin by briefly explaining how trace relations may be 
proved in general terms from the known resolvents of these theories.
Then, we discuss the derivation of trace relations from the explicit localization
computation of the chiral observables $\T{n}$.

\subsection{Trace relations from resolvent expansion}

In the pure gauge theory, trace relations may be systematically obtained from the 
 resolvent for chiral 1-point functions   \cite{Cachazo:2002ry}
\be
\la{2.1}
\vev{\tr\frac{1}{z-\varphi}} = \frac{2\,z}{\sqrt{\phantom{\frac{.}{.}}(z^{2}-e^{2})^{2}-4\,q}\ },
\ee
where $q$ is the conventional instanton counting parameter. The coordinate $e$ may be 
conveniently traded by the moduli space coordinate $\bm{u}=\T{2}$.
\footnote{Here, we use a boldface symbol for the scalar $\T{2}$
just to emphasize  it better in the 
following equations.}
Expanding at large $z$ and comparing the two sides of (\ref{2.1})
%\be
%\la{2.2}
%\frac{2}{x}+\frac{\vev{\tr\varphi^{2}}}{x^{3}}+\frac{\vev{\tr\varphi^{4}}}{x^{5}}+\dots=
%\frac{2}{x}+\frac{2e^{2}}{x^{3}}+\frac{2e^{4}+4q}{x^{5}}+\dots, 
%\ee
we immediately recover the well-known relations
\begin{align}
\la{2.2}
\T{2} &= \bm{u}, &  \T{4}  &= \frac{1}{2}\,\bm{u}^{2}+4\,q, \notag \\
\T{6} &= \frac{1}{4}\,\bm{u}^{3}+6\,q\,\bm{u}, & 
\T{8} &= \frac{1}{8}\,\bm{u}^{4}+6\,q\,\bm{u}^{2}+12\,q^{2}, \qquad \text{\em etc.}
\end{align}
The same strategy may be applied to the $\mc N=2^{\star}$ theory as in the analysis of 
\cite{Fucito:2005wc}. For completeness, we briefly review the construction. The main 
tool is the D'Hoker-Phong formulation of the spectral curve \cite{D'Hoker:1997ha}. 
In the $SU(2)$ theory, 
%For the $SU(2)$ gauge group, it  reads 
%\be
%\la{2.3}
%x^{2}-e^{2}+2\,m\,x\frac{\vartheta_{1}'(z)}{\vartheta_{1}(z)}+
%m^{2}\,\frac{\vartheta_{1}''(z)}{\vartheta_{1}(z)}=0.
%\ee
following  \cite{Fucito:2005wc,Nekrasov:2003rj} to which we defer the reader for 
a thorough discussion, one can introduce the resolvent
\be
\label{2.4}
G(z) = \vev{\tr\frac{1}{z-\varphi-\frac{i}{2}\,m_{c}}}
- \vev{\tr\frac{1}{z-\varphi+\frac{i}{2}\,m_{c}}},
\ee
and aims at a generalization of the nice formula (\ref{2.1}). The resolvent $G(z)$
 is expected to be analytic with branch cuts 
$[\alpha_{n}^{-}\pm\frac{i}{2}\,m_{c}, \alpha_{n}^{+}\pm\frac{i}{2}\,m_{c}]$, $n=1,2$.
The function 
$\omega(z) = \frac{1}{2\,\pi\,i}\int_{\infty}^{z}G(y)\,dy$
maps the cut complex plane to the curve $(\omega(z), z)$ which is a double cover of an elliptic curve
obtained by the identification ($\pi$ factor omitted)
$(\omega+1,z)\sim(\omega,z)$ and $(\omega+\tau,z)\sim(\omega,z+i\,m_{c})$ \cite{Kazakov:1998ji}.
This is the double periodicity of the function \cite{D'Hoker:1997ha}
$
f(\omega,z) = z^{2}-2\,i\,m_{c}\,z\,h_{1}(\omega)+(i\,m_{c})^{2}\,h_{2}(\omega)-z_{1}^{2}
$
where
\be
\label{2.5}
h_{1}(\omega) = \frac{\vartheta_{1}'(\omega|\tau)}{\vartheta_{1}(\omega|\tau)}, \qquad
h_{2}(\omega) =  \frac{\vartheta_{1}''(\omega|\tau)}{\vartheta_{1}(\omega|\tau)}
= h_{1}'(\omega)+h_{1}^{2}(\omega)
\ee
The resolvent is obtained as the large $z$ expansion of 
$G(z)= 2\,\pi\,i\,\omega'(z) $ where $\omega(z)$ is implicitly defined by 
\be
\label{2.6}
f(\omega(z), -2\,\pi\,i\,z)=0.
\ee
This can be solved perturbatively at small $\omega\sim 1/z$. \footnote{
\la{f6} 
This is 
achieved by using the representation 
\be
\notag 
h_{1}(\omega) = \pi\,\cot(\pi\,\omega)+4\,\pi\,\sum_{n=1}^{\infty}\frac{q^{n}}{1-q^{n}}
\sin(2\,\pi\,n\,\omega) = \frac{1}{\omega}+\mc O(\omega).
\ee
Higher orders in $\omega$  involve the Eisenstein functions, see App.~(\ref{app:eisen}), since, for instance,
$\E_{2} = 1-24\,f_{1}$, $\E_{4}=1+240\,f_{3}$, and $\E_{6}=1-504\,f_{5}$
where $f_{p}=\sum_{n=1}^{\infty} \frac{n^{p}\,q^{n}}{1-q^{n}}$.}
%The relevant solution of (\ref{2.6}) is the one singular in the limit $z\to 0$, {\em i.e.}
%the branch $\omega = -\frac{m_{c}}{\pi\,z}+\mc O(z^{-3})$.
%$z=-m_{c}/(\pi\,\omega)+\mc O(\omega)$.
%\be
%\label{A.6}
%z = \frac{1}{2\,\pi}\bigg(-m_{c}\,h_{1}-\sqrt{-m_{c}^{2}\,h_{1}'-z_{1}^{2}}\bigg)
%= -\frac{m_{c}}{\pi\,\omega}+\mc O(\omega).
%\ee
%This can be inverted to write $\omega$ as a series in $1/z$ with leading term 
%$\omega = -\frac{m_{c}}{\pi\,z}+\mc O(z^{-3})$.
Comparing $G(z)= 2\,\pi\,i\,\omega'(z) $ with the large $z$ expansion of (\ref{2.4})
%\be
%\label{2.8}
%G(z) 
%= \frac{2\,i\,m_{c}}{z^{2}}+\frac{1}{z^{4}}\,\bigg(
%\frac{(i m_{c})^{3}}{2}+3im_{c}\T{2}
%\bigg)
%+\frac{1}{z^{6}}\,\bigg(
%\frac{(im_{c})^{5}}{8}+\frac{5}{2}(im_{c})^{3}\T{2}+5im_{c}\T{4}
%\bigg)+\dots\ ,
%\ee
we obtain 
\begin{align}
\label{2.7}
\T{2} &= 4\,f_{1}\,m_{c}^{2}-\frac{z_{1}^{2}}{2\pi^{2}}, \notag \\
\T{4} &= \frac{1}{2}(\T{2})^{2}+8m_{c}^{2}f_{1}\T{2}+m_{c}^{4}\bigg(
\frac{4}{3}f_{1}-32f_{1}^{2}+\frac{8}{3}f_{3}\bigg),
\end{align}
and so on. Using the first equation to replace $z_{1}$ by $\bm{u}=\T{2}$ and setting $im_{c}=m$ gives 
all the desired trace relations. The first cases are
\begin{align}
\la{2.8}
\T{4} &=  \frac{\bm{u}^{2}}{2}+\frac{1}{3} (\,\E_2-1) m^2 \,\bm{u}+\frac{1}{90}
   m^4 \left(-5 \,\E_2^2+5 \,\E_2+\,\E_4-1\right), \notag \\
\T{6} = & \frac{\bm{u}^3}{4}+\frac{1}{2} (\,\E_2-1) m^2\,\bm{u}^{2}+\frac{1}{4}
   m^4 (1-\,\E_2)\,\bm{u}+\notag  \\ 
   & \frac{m^6 \left(-140 \,\E_2^3+525
   \,\E_2^2+84 \,\E_2 (\,\E_4-5)-105 \,\E_4-24
   \,\E_6+80\right)}{7560}\notag, \\
   \T{8} &= \frac{\bm{u}^{4}}{8} + \frac{1}{2} \left( \E_2  -1\right) m^2\,\bm{u}^{3} + \notag\\
& \left( \frac{7}{36} \E_2^2 - \frac{11}{12} \E_2 - \frac{1}{36} \E_4 + \frac{3}{4}  \right) m^4 
\,\bm{u}^2 + \notag \\	
& \left(- \frac{1}{18} \E_2^3 + \frac{7}{108} \E_2^2 
+ \frac{1}{3} \E_2 + \frac{7}{270} \,\E_4 \,\E_2 - \frac{1}{108} \E_4 
- \frac{1}{135} \E_6 -\frac{19}{54} \right) m^6\,\bm{u} +\notag \\
&\left( -\frac{\E_2^4}{648} + \frac{11 \, \E_2^3}{324} + \frac{\E_4 \, \E_2^2}{324}  - \frac{35\,\E_2^2}{324} - \frac{7\E_2 \, \E_4}{324} - \frac{\E_6 \, \E_2}{567} + \frac{\E_2}{12} +
\frac{\E_4^2}{4536} + \right. \notag\\ 
&\qquad \left.  \frac{71 \E_4}{3240} + \frac{\E_6}{162} - \frac{17}{1080} \right) m^8.
\end{align}
These are non-trivial exact all-instanton relations. It is remarkable that they take such a simple form 
even though the separate $\T{n}$ are definitely non trivial and, for instance, require  
an infinite series of corrections in the small mass $m$ expansion.

\subsection{Trace relations from localization}

There is a simple trick to generate relations like  (\ref{2.2})  \cite{Fucito:2005wc}.
To illustrate it, we start from the explicit expression of $\T{n}$ computed by localization methods that
we briefly review in App.~(\ref{app:localization}). 
In the  $\mc N=2$ pure gauge theory the explicit values of $\T{n}$  take 
the form of an equivalent expansion in instanton number or large scalar field vacuum
expectation value $a$. They read 
\begin{align}
\la{2.9}
\T{2} &= 2 \,a^2+\frac{q}{a^2}+\frac{5\, q^2}{16\, a^6}+\frac{9 \,q^3}{32
   \,a^{10}}+\frac{1469 \,q^4}{4096 \,a^{14}}+\frac{4471
   \,q^5}{8192 \,a^{18}}+\dots, \notag \\
\T{4} &= 2 \,a^4+6 \,q+\frac{9 \,q^2}{8 \,a^4}+\frac{7 \,q^3}{8
   \,a^8}+\frac{2145 \,q^4}{2048 \,a^{12}}+\frac{1575
   \,q^5}{1024 \,a^{16}}+\dots, \notag \\
\T{6} &= 2 \,a^6+15 \,a^2 \,q+\frac{135 \,q^2}{16 \,a^2}+\frac{125 \,q^3}{32
   \,a^6}+\frac{16335 \,q^4}{4096 \,a^{10}}+\frac{44343
   \,q^5}{8192 \,a^{14}}+\dots, \notag \\
\T{8} &= 2 \,a^8+28 \,a^4 \,q+\frac{161 \,q^2}{4}+\frac{35 \,q^3}{2
   \,a^4}+\frac{15337 \,q^4}{1024 \,a^8}+\frac{19173 \,q^5}{1024
   \,a^{12}}+\dots\ ,
\end{align}
with vanishing odd traces. We can use the first of these series expansion to 
write $a = a(\bm{u})$ 
with $\bm{u}\equiv \T{2}$. Replacing in the other equations we get 
for the pure gauge theory the  (exact) relations (\ref{2.2}). A similar procedure may be 
applied to the  $\mc N=2^{\star}$ theory. A closed form cannot be obtained, but 
the full $q$ dependence can be resummed. This is achieved by means of 
an educated Ansatz taking into account the modular properties encoding $S$-duality
 \cite{Minahan:1997if,Billo:2013fi,Billo:2013jba,Ashok:2016ewb}. 
Organizing $\T{n}$ as a mass expansion, 
the results are,  see also \cite{Ashok:2016ewb} \footnote{
Notice that $\T{n}$ with $n\ge 6$ differs from what is obtained using (5.11) of \cite{Ashok:2016oyh}, the discrepancy starting at order $\sim m^{6}\,q^{2}$. This is due to the fact that the compact formulas
in \cite{Ashok:2016oyh} are explicitly designed to interpolate the $n\le 5$ cases. We thank A. Lerda for 
clarifications on this issue.}
\begin{align}
\label{2.10}
\T{2} &=  2 a^2+\frac{1}{6} \left(\E_2-1\right) m^2+\frac{\left(\E_4-\E_2^2\right) m^4}{288 a^2}+\frac{\left(-5 \E_2^3+3 \E_4 \E_2+2 \E_6\right) m^6}{17280 a^4}+ \mc O\left(m^8\right), \notag \\
\T{4} &=  2 a^4+ a^2 \left(\E_2-1\right) m^2+\frac{1}{720}
 \left(5 \E_2^2-60 \E_2+13 \E_4+42\right) m^4+
\notag\\ &  \frac{\left(-20 \E_2^3+15 \E_2^2+18 \E_4 \E_2-15 \E_4+2 \E_6\right) }{8640 a^2}\,m^6+
 \mc O(m^8).\notag \\
 \T{6} &=  2 a^6+\frac{5}{2} a^4 \left(\E_2-1\right) m^2+\frac{1}{96} a^2 \left(35 \E_2^2-120 \E_2+\E_4+84\right) m^4+
\notag\\ 
& \frac{\left(-525 \E_2^3-350 \E_2^2+21 \left(39 \E_4+140\right) \E_2-2 
\left(455 \E_4+57 \E_6+930\right)\right)
  }{40320}\,  m^6 + \mc O(m^8), \notag \\
\T{8} &=  2 a^8 + \frac{14}{3} a^6 \left(\E_2-1\right) m^2 + \frac{a^4}{72}\left( 294 - 420 \E_2 + 133 \E_2^2 - 7 \E_4 \right)\, m^4 +
\notag\\ &  \frac{a^2}{4320} \left( -5580 + 8820 \E_2 - 3675 \E_2^2  - 105 \E_4 
+ 350 \E_2^3 + 252 \E_2 \E_4 - 62 \E_6 \right) m^6 + \mc O(m^8),
\end{align}
where $\E_{n}(q)$ are the Eisenstein series defined in App.~(\ref{app:eisen}). 
Repeating the trick of writing $a=a(\bm{u})$ starting from these expressions, we arrive
at the previous trace relations in (\ref{2.8}).

\section{Gravitational corrections to trace relations in pure $SU(2)$ $\mc N=2$ theory}
\la{sec:gravpure}

We now move to the more interesting case of $\eps$-deformed pure $SU(2)$ $\mc N=2$ theory
where we look for a generalization of the relations (\ref{2.2}). Since the $\eps$-deformed
resolvent is not available, we try to work out such relations from the explicit localization
results. 
On a generic $\Omega$-background, the values of  $\T{n}$ from localization 
may be organized again in a large $a$ expansion, but the expressions for generic 
 $a$, $\eps_{1}$, $\eps_{2}$ are quite complicated. Just to illustrate some of the results, 
we show  the expansion around the undeformed limit $\bm{\eps}=0$ at third order
for $\T{2}$
\begin{align}
\la{uuu1}
\T{2} =\, & 
2 \,a^2+\frac{q}{a^2}+\frac{5 \,q^2}{16 \,a^6}+\frac{9 \,q^3}{32 \,a^{10}}
+\frac{1469 \,q^4}{4096 \,a^{14}}+ \dots\notag 
\\ \notag & +
\eps_1 \eps_2 \left(-\frac{q^2}{8 \,a^8}-\frac{q^3}{2 \,a^{12}}
-\frac{1647 \,q^4}{1024 \,a^{16}}+ \dots\right)
\\ \notag & +
(\eps_1+\eps_2)^2 \left(\frac{q}{4 \,a^4}+\frac{21 \,q^2}{32 \,a^8}
+\frac{55 \,q^3}{32\,a^{12}}+\frac{18445 \,q^4}{4096 \,a^{16}}+ \dots\right)
\\ \notag & +
(\eps_1 \eps_2)^2 \left(\frac{11 \,q^2}{256 \,a^{10}}+\frac{351 \,q^3}{512 \,a^{14}}
+\frac{171201 \,q^4}{32768 \,a^{18}}+ \dots\right) 
\\ \notag & +
\eps_1 \eps_2 (\eps_1+\eps_2)^2 \left(-\frac{35 \,q^2}{64\,a^{10}}
-\frac{689 \,q^3}{128 \,a^{14}}
-\frac{269693 \,q^4}{8192 \,a^{18}}+ \dots\right) 
\\ \notag & +
(\eps_1 \eps_2)^3 \left(-\frac{7 \,q^2}{512 \,a^{12}}-\frac{879 \,q^3}{1024 \,a^{16}}
-\frac{985823 \,q^4}{65536 \,a^{20}}+ \dots\right)
\\  & +
(\eps_1 \eps_2)^2 (\eps_1+\eps_2)^2 \left(\frac{325 \,q^2}{1024\,a^{12}}
+\frac{23631 \,q^3}{2048 \,a^{16}}+\frac{20930787 \,q^4}{131072 \,a^{20}}+ \dots\right) + \dots
\end{align}
Similar expansions for higher order traces are collected in App.~(\ref{app:pure}), but we stress 
again that we shall always work with the exact localization expressions 
of $\T{n}$, {\em i.e.} not using the small $\eps_{1}, \eps_{2}$ expansion in (\ref{uuu1})
that was just a device for illustration.

\subsection{Surprises from empirical trace relations}
\la{sec:empirical-su2}

To find trace relations in this case, 
we can try to repeat the practical procedure that worked in the undeformed theories.
In other words, we invert the relation between
$a$ and $\T{2}$ and replace in the higher traces $\T{n}$. Doing so, we don't find 
a  closed expression generalizing (\ref{2.2}).
Nevertheless, we have been able to propose the following relations that we checked 
at the level of 10 instantons and, of course, without expansion in $\bm{\eps}$.
Let us denote
\be
\bm{u} = \T{2}, \qquad X' = q\,\partial_{q}\,X, \qquad X'' = (q\,\partial_{q})^{2}\,X, \ 
\qquad
\text{\em etc.}
\ee
Then, we find the following results
\begin{align}
\la{3.3}
\T{3} &= 0, \notag \\
\T{4} &= \frac{1}{2}\,\bm{u}^{2}+4\,q-\eps_{1}\,\eps_{2}\,\bm{u}', \notag \\
\T{5} &= 10\,(\eps_{1}+\eps_{2})\,q, \notag \\
\T{6} &= \frac{1}{4}\,\bm{u}^{3}+6\,q\,\bm{u}-\frac{3}{2}\,\eps_{1}\,\eps_{2}\,\bm{u}\,\bm{u}'+
\eps_{1}^{2}\eps_{2}^{2}\,\bm{u}''+6\,q\,(3\,\eps_{1}^{2}+
4\,\eps_{1}\,\eps_{2}+3\,\eps_{2}^{2}), \notag \\
\T{7} &= (\eps_{1}+\eps_{2})\,\bigg[
21\,q\,\bm{u}+7\,q\,(4\,\eps_{1}^{2}+3\,\eps_{1}\,\eps_{2}+4\,\eps_{2}^{2})\bigg], \notag \\
\T{8} &= \frac{1}{8}	\,\bm{u}^{4}+6\,q\,\bm{u}^{2}+12\,q^{2}-\eps_{1}^{3}\eps_{2}^{3}\,\bm{u}'''
+2\,\eps_{1}^{2}\eps_{2}^{2}\,\bm{u}\,\bm{u}''+\frac{3}{2}\,\eps_{1}^{2}\eps_{2}^{2}\,
\bm{u}^{\prime\,2}-12\,\eps_{1}\eps_{2}\,q\,\bm{u}' \notag \\
&-\frac{3}{2}\,\eps_{1}\,\eps_{2}\,\bm{u}^{2}\,\bm{u}'+(52\,\eps_{1}^{2}+72\,\eps_{1}\eps_{2}
+52\,\eps_{2}^{2})\,q\,\bm{u}\notag \\
&+8\,q\,(5\,\eps_{1}^{4}+11\,\eps_{1}^{3}\eps_{2}
+15\eps_{1}^{2}\eps_{2}^{2}+11\eps_{1}\eps_{2}^{3}+5\eps_{2}^{4}). 
\end{align}
These expressions show clearly the reason why the naive procedure of replacing $a$ as
a function of $\T{2}$ did not work.
In (\ref{3.3}) there are non trivial derivatives of $\bm{u}$, the r.h.s's are not polynomials in 
$\bm{u}$ when $\eps_{1}\eps_{2}\neq 0$. We also remark that 
the relations (\ref{3.3}) simplify, but do not trivialize in the  Nekrasov-Shatashvili (NS) limit
\be
\la{3.4}
\eps_{1}=\hbar, \qquad \eps_{2}=0.
\ee
In fact, 
all derivative terms vanish in this case and the trace relations read \footnote{
Similar corrections have been investigated in \cite{Fujii:2007qe} in the limit
$\eps_{1}=-\eps_{2}$ and in the $U(1)$ gauge theory.
}
\begin{align}
\la{3.5}
\T{3} &= 0,     & \T{4} &= \frac{1}{2}\,\bm{u}^{2}+4\,q,  \notag \\
\T{5} &= 10\,\hbar\,q,     & \T{6} &= \frac{1}{4}\,\bm{u}^{3}+6\,q\,\bm{u}
+18\,\hbar^{2}\,q, \notag \\
\T{7} &=\hbar\,(
21\,q\,\bm{u}+28\,\hbar^{2}\,q), & 
\T{8} &= \frac{1}{8}	\,\bm{u}^{4}+6\,q\,\bm{u}^{2}+12\,q^{2}
+52\,\hbar^{2}\,q\,\bm{u}+40\,\hbar^{4}\,q. 
\end{align}
We can  show how the peculiar $\hbar$ dependent corrections in (\ref{3.5})
and deforming the previous (\ref{2.2})
can be predicted from the 
deformed SW curve of pure gauge $SU(2)$ theory. This is discussed in App.~(\ref{app:pogho}).

\section{Gravitational corrections to trace relations in $\mc N=2^{\star}$. The NS limit}
\la{sec:star-NS}

We begin the analysis of the $\mc N=2^{\star}$ theory in the Nekrasov-Shatashvili limit. Indeed, 
given our experience in the pure gauge theory, we expect major simplifications to occur 
when one of the deformation parameters vanishes. Later, we shall study the case of a 
fully deformed background.

\subsection{Localization results}

We can compute the chiral traces at some high instanton order from localization and 
then take the NS limit. The explicit expressions of $\T{n}$ are rather involved. To illustrate
them, we  trade the hypermultiplet
mass by the following combination 
\be
\Chbar = 4 m^2-\hbar^{2}, \qquad \hbar \equiv \eps_{1}.
\ee
Then, for $\T{2}$ we obtain 
\begin{align}
\la{4.2}
& \T{2} = 2\,a^{2}+\Chbar\,\bigg(\frac{\Chbar}{16 a^2-4 \hbar ^2}-1\bigg)\,q+
\Chbar\,\bigg[
\frac{\Chbar^3 (20 a^2+7 \hbar ^2)}{256 (a^2-\hbar ^2)(4 a^2-\hbar
   ^2)^3}\notag \\
   &-\frac{3 \Chbar^2}{16 (a^2-\hbar ^2) (4 a^2-\hbar
   ^2)}+\frac{3 \Chbar (2 a^2-\hbar ^2)}{4 (a^2-\hbar ^2)
   (4 a^2-\hbar ^2)}-3 
\bigg]\,q^{2}\notag \\
&+\Chbar\,\bigg[
\frac{\Chbar^{5} (144 a^4+232 a^2 \hbar ^2+29 \hbar ^4)}{512 (4 a^2-9 \hbar
   ^2) (a^2-\hbar ^2) (4 a^2-\hbar ^2)^5} -\frac{\Chbar^{4} (28
   a^2+17 \hbar ^2)}{32 (4 a^2-9 \hbar ^2) (a^2-\hbar
   ^2) (4 a^2-\hbar ^2)^3}\notag \\
   &+\frac{\Chbar^{3} (120 a^4-74 a^2 \hbar
   ^2-\hbar ^4)}{8 (4 a^2-9 \hbar ^2) (a^2-\hbar ^2) (4
   a^2-\hbar ^2)^3}-\frac{6 \Chbar^{2}}{(4 a^2-9 \hbar ^2) (4 a^2-\hbar
   ^2)}\notag \\
   & +\frac{3 \Chbar (4 a^2-3 \hbar ^2)}{(4 a^2-9 \hbar ^2)
   (4 a^2-\hbar ^2)}-4 
   \bigg]\,q^{3}+\mc O(q^{4}).
\end{align}
The same expansion for $\T{3}$ is much simpler
\be
\la{4.3}
\T{3} = \Chbar\,\hbar\,\bigg(-\frac{3}{2}\,q-\frac{15}{2}
\,q^{2}-15  \,q^{3}+\mc O(q^{4})\bigg),
\ee
and clearly vanish for $\hbar\to 0$. Besides, it is independent on $a$. This will be false for the
higher odd traces. The next even trace is similar to (\ref{4.2}) and reads
\begin{align}
\la{4.4}
\T{4} &=2\,a^{4}+\Chbar\,\bigg[\frac{1}{4} \Chbar \bigg(1-\frac{2 a^2}{\hbar ^2-4 a^2}\bigg)
-2 (3 a^2+\hbar
   ^2)\bigg]\,q+\Chbar\,\bigg[
   \frac{\Chbar^3 (36 a^4-13 a^2 \hbar ^2+4 \hbar ^4)}{128 (a^2-\hbar ^2)
   (4 a^2-\hbar ^2)^3}\notag \\
   &-\frac{3 \Chbar^2 (3 a^2-2 \hbar ^2)}{8
   ((a^2-\hbar ^2) (4 a^2-\hbar ^2))}+\frac{3 \Chbar (16
   a^4-17 a^2 \hbar ^2+3 \hbar ^4)}{4 (a^2-\hbar ^2) (4
   a^2-\hbar ^2)}-18  (a^2+\hbar
   ^2)
   \bigg]\,q^{2}\notag \\
&   +\Chbar\,\bigg[
   \frac{\Chbar^{5} (896 a^6+240 a^4 \hbar ^2+16 a^2 \hbar ^4+63 \hbar ^6)}{1024
   (4 a^2-9 \hbar ^2) (a^2-\hbar ^2) (4 a^2-\hbar
   ^2)^5}-\frac{\Chbar^{4} (880 a^4-664 a^2 \hbar ^2-81 \hbar ^4)}{256
   ((4 a^2-9 \hbar ^2) (a^2-\hbar ^2) (4 a^2-\hbar
   ^2)^3)}\notag \\
   &+\frac{\Chbar^3 (576 a^6-1084 a^4 \hbar ^2+427 a^2 \hbar ^4-54
   \hbar ^6)}{8 (4 a^2-9 \hbar ^2) (a^2-\hbar ^2) (4
   a^2-\hbar ^2)^3}-\frac{3 \Chbar^2 (52 a^4-121 a^2 \hbar ^2+54 \hbar ^4)}{4
   ((4 a^2-9 \hbar ^2) (a^2-\hbar ^2) (4 a^2-\hbar
   ^2))}\notag \\
   &+\frac{2 \Chbar (76 a^4-169 a^2 \hbar ^2+36 \hbar ^4)}{(4
   a^2-9 \hbar ^2) (4 a^2-\hbar ^2)}-8  (3 a^2+7 \hbar
   ^2)
      \bigg]\,q^{3}+\mc O(q^{4}), 
\end{align}
and so on. We can invert the relation (\ref{4.2}) to express 
\be
a = a(\bm{u}), \qquad \bm{u}\equiv \T{2},
\ee
order by order in the instanton number. Then, we replace $a(\bm{u})$ in (\ref{4.4}) and obtain
the following quite simple expansions where the dependence on $\bm{u}$ turns out to be {
simply polynomial} (we add a few more chiral traces)
\begin{align}
\la{4.6}
\T{4} &= \frac{1}{2}\,\bm{u}^{2}+(
-2 q-6 q^2-8 q^3+\mc O(q^4))\,\Chbar\,\bm{u} \notag \\
&+\frac{1}{4} \Chbar\, q (\Chbar-8 \hbar ^2)-\frac{1}{4} q^2 \Chbar (\Chbar
+72 \hbar
   ^2)-7 q^3 \Chbar (\Chbar+8 \hbar ^2)+\mc O(q^{4}), \notag \\
\T{5} &= \bigg(
-\frac{15 q  }{2}-\frac{75 q^2  }{2}-75 q^3  +\mc O(q^{4})\bigg)\,\hbar\,\Chbar\,\bm{u}
\notag \\
&+\bigg(\frac{5}{8} q (\Chbar-4 \hbar^{2} )+\frac{5}{8} q^2 \left(3 \Chbar-68 \hbar
   ^2\right)-\frac{5}{4} q^3 \left(11 \Chbar+164 \hbar ^2\right)+\mc O(q^{4})\bigg)\,\Chbar\,\hbar
   , \notag \\
\T{6} &= \frac{1}{4}\,\bm{u}^{3}+(
-3 q-9 q^2-12 q^3+\mc O(q^4))\,\Chbar\,\bm{u}^{2}\notag\\
&+\bigg(
q \left(\frac{3 \Chbar}{8}-15 \hbar ^2\right)+q^2 \left(\frac{9 \Chbar}{8}-135 \hbar ^2\right)+q^3
   \left(\frac{3 \Chbar}{2}-420 \hbar ^2\right)
\bigg)\,\Chbar\,\bm{u}\notag \\
&+\Chbar\,\bigg(
\frac{3}{8} q \hbar ^2 \left(3 \Chbar-8 \hbar ^2\right)-\frac{3}{16} q^2 \left(\Chbar^2-78 \Chbar
   \hbar ^2+528 \hbar ^4\right) \\ \notag
   &-\frac{3}{2} q^3 \left(\Chbar^2-18 \Chbar \hbar ^2+488 \hbar
   ^4\right)+\mc O(q^4) 
   \bigg).
\end{align}
We stress that we 
have obtained the relations  (\ref{4.6}) from explicit localization computations, see the explicit
results in (\ref{4.2}-\ref{4.4}). \footnote{These must be worked out at generic $\eps_{1}\equiv\hbar$, 
and with non zero $\eps_{2}$.
Only at the end we can take the $\eps_{2}\to 0$ limit.}

\medskip
As a cross-check of our calculations
it is interesting to consider the proposal in \cite{Fucito:2011pn}. The authors of this paper
analyze the NS limit of the Nekrasov integrals by saddle point methods. {The advantage is 
that they are able to work directly at $\eps_{2}=0$}.
We apply the results of  \cite{Fucito:2011pn} to the $SU(2)$ $\mc N=2^{\star}$ case
in App.~(\ref{app:fucito2}) with full agreement.

\subsection{Empirical  all-instanton  trace relations in the NS limit}

With some educated guess, it is possible to identify the power series in $q$ in terms of Eisenstein series
and {their odd generalizations}, see App.~(\ref{app:eisen}). The results up to $\T{7}$ are summarized in the following expressions
\begin{align}
\la{4.7}
\T{3} =\, & -\frac{3}{2} \,\Chbar \,\E_3 \hbar  , \notag \\
\T{4} =\, & \frac{\bm{u}^2}{2}+\frac{1}{12} \bm{u} \,\Chbar (\,\E_2-1)
-\frac{\,\Chbar \left(\,\Chbar \left(5 \,\E_2^2-5 \,\E_2-\,\E_4+1\right)+12 (\,\E_4-1) \hbar ^2\right)}{1440} , \notag \\
\T{5} =\, & -\frac{15}{2} \bm{u} \,\Chbar \,\E_3 \hbar  + \frac{5}{8} 
\,\Chbar \hbar  \left(2 \,\Chbar \,\E_3'-\,\Chbar \,\E_5-4 \,\E_5 \hbar ^2\right), \notag \\
\T{6} =\, & \frac{\bm{u}^3}{4}+\frac{1}{8} \bm{u}^2 \,\Chbar (\,\E_2-1)-\frac{1}{64} \bm{u} \,\Chbar \left(-\,\Chbar+\,\Chbar \,\E_2+4 \,\E_4 \hbar ^2-4 \hbar ^2\right)
  \notag \\ \quad &
-\frac{\,\Chbar^3 \left(140 \,\E_2^3-525 \,\E_2^2-84 \,\E_2 (\,\E_4-5)+105\,\E_4+24 \,\E_6-80\right)}{483840}
  \notag  \\ \quad &
+\frac{\Chbar^{2}\hbar^{2}(7\E_{2}\E_{4}-2\E_{6}+25200\E_{3}^{2}-5}{2240}
+\frac{1}{168}\,\Chbar\,\hbar^{4}\,(\E_{6}-1) , \notag \\
\T{7} =\, & -\frac{105}{8} \bm{u}^2 \,\Chbar \,\E_3 \hbar  +  \frac{7}{16} 
\bm{u} \,\Chbar \hbar  \left(18 \,\Chbar \,\E_3'-5 \,\Chbar \,\E_2 \,\E_3
+5 \,\Chbar \,\E_3-15 \,\Chbar \,\E_5-60 \,\E_5 \hbar ^2\right) \notag \\ 
\quad & +\frac{7}{384}\Chbar^{3}\,\hbar\,\bigg[(5\E_{2}^{2}-5\E_{2}-\E_{4}+1)\E_{3}
-12\E_{3}'-24 \E_{3}''+48 \E_{5}'-12\E_{7}\bigg] \\ \notag
\quad & +\frac{7}{32}\Chbar^{2}\hbar^{3}\bigg[(\E_{4}-1)\E_{3}+16\E_{5}'-8\E_{7}\bigg]
-\frac{7}{2}\Chbar\,\hbar^{5}\E_{7}. 
\end{align}
The relations (\ref{4.7}) are quite interesting because they are valid at all instanton numbers.
They are the $\mc N=2^{\star}$ version of the much simpler relations (\ref{3.5})
valid in the pure gauge theory. To prove them in a systematic way, 
we would need a {deformed version of D'Hoker-Phong curve} with the 
full $q$ dependence packaged in Eisenstein series or related objects.
Unfortunately, this is not available.

\section{Trace relations in $\mc N=2^{\star}$ on a generic $\Omega$-background}
\la{sec:star-full}

One can look for trace relations in the generic background with 
non zero $\eps_{1}$, $\eps_{2}$ parameters.
Given our experience in the pure gauge theory, we expect these relations to involve derivatives
of the moduli space coordinate $\bm{u} = \T{2}$ making them highly non-trivial.
With some insight, we have been able to find them, testing always at high ($\ge 10$)
explicit localization results. 

\subsection{Empirical trace relations for generic $\Omega$-background}
\label{sec:empirical}

We denote as always $\bm{u} = \T{2}$ and introduce the notations
\be
\la{zzz1}
X' \equiv q\,\frac{d}{dq} X, %\qquad \T{n}_{0} = \left. \T{n}\right|_{\bm{\eps}=0}, \\
\qquad \mathcal{C} = 4 m^2-(\eps_1-\eps_2)^2, \qquad \p=\eps_{1}\,\eps_{2}, \qquad
\s=\eps_{1}+\eps_{2}.
\ee
The explicit instanton expansion of $\T{n}$ is highly non-trivial at generic finite $a,m,\eps_{i}$. Nevertheless,
we found the following relations.

\medskip
\noindent
\underline{$n=3$}

\medskip
\noindent
This odd trace is computed by  
\be
\T{3} = -\frac{3}{2}\,\mc C\,\s\, \E_3.
\ee
In general, all odd traces must vanish in the undeformed limit $\eps_{i}\to 0$. In our calculation, 
this will be always due to an explicit $\s=\eps_{1}+\eps_{2}$ prefactor, see also $\T{5}$ below.

\medskip
\noindent
\underline{$n=4$}

\medskip
\noindent
The first non trivial even trace is given by the compact expression
\begin{align}
\la{5.3}
\T{4} &= \frac{1}{12} \mathcal{C} \left(\,\E_2-1\right) \,\bm{u}
- \p\,  \bm{u}'  +\frac{\,\bm{u}^2}{2} + N_4,
\end{align}
where $N_{4}$ does not depend on $a$ and is given by 
\begin{align}
N_4 = -\frac{\mathcal{C}^2}{1440}-\frac{1}{120} \mathcal{C} \left(\,\p-\,\s^2\right)
-\frac{1}{288} \mathcal{C}  (\mathcal{C}-3 \,\p)\,\E_2^2+\frac{\mathcal{C}^2 \,\E_2}{288}
+\frac{\mathcal{C}  \left(\mathcal{C}-3 \left(\,\p+4 \,\s^2\right)\right)}{1440}\,\E_4.
\end{align}
The undeformed and NS limits of this relation reproduce (\ref{2.8}) and (\ref{4.7}), respectively.
We checked this for the higher traces too finding always agreement. As we expected, there is 
a contribution $\sim \bm{u}'$ whenever $\eps_{1}\eps_{2}\neq 0$.

\medskip
\noindent
\underline{$n=5$}

\medskip
\noindent
In this case, we found
\begin{align}
\la{5.5}
\T{5} =\, & -\frac{15}{2} \, \mathcal{C}  \,\s\,\E_3\, \bm{u}
+\frac{5}{4} \,\mathcal{C} \,\s\,(\mc C-4\,\p)\,\E_3' 
-\frac{5}{8} \mathcal{C}  \,\s\, \left(\mathcal{C}-8 \,\p+4 \,\s^2\right)\,\E_5.
\end{align}

\medskip
\noindent
\underline{$n=6$}

\medskip
\noindent

\begin{align}
\la{5.6}
\T{6} =\, & \frac{\bm{u}^3}{4}+\frac{1}{8} \mathcal{C} (\,\E_2-1)\, \bm{u}^2
-\frac{3}{2}\,\p\, \bm{u} \,\bm{u}' -\frac{1}{4} \, \mathcal{C}\,\p\, (\,\E_2-1) \,\bm{u}'
+\p^{2}\,\bm{u}'' 
\notag \\  &  
-\frac{1}{192} \, \mathcal{C} \left(\,\p \left(-7 \,\E_2^2-5 \,\E_4+12\right)
+3 \mathcal{C} (\,\E_2-1)
+12 \,\s^{2}\,(\,\E_4-1) \right)\,\bm{u}+N_{6},
\end{align}
where $N_{6}$ is the following combination independent on $a$
\begin{align}
N_6 =\, & \frac{\mathcal{C}^3}{6048}+\frac{\mathcal{C}^2 \left(2 \,\p-3 \,\s^2\right)}{1344}
-\frac{1}{168} \mathcal{C} \left(\,\p^2-3 \,\p \,\s^2
+\,\s^4\right)-\frac{\mathcal{C} \, (\mathcal{C}-3 \,\p) (\mathcal{C}+2 \,\p)}{3456}\,\E_2^3
 \\  &  
+\frac{\mathcal{C}^2 \,(5 \mathcal{C}-8 \,\p)}{4608}\,\E_2^2 
+\frac{\mathcal{C} \,\E_2 \left(\mathcal{C}^2 (\,\E_4-5)+\mathcal{C} 
\,\E_4 \left(18 \,\s^2-17 \,\p\right)+6 \,\E_4 \,\p \left(7 \,\p-12 \,\s^2\right)\right)}{5760}
\notag \\  &  
+\frac{45}{4} \mathcal{C}^2 \,\s^2\,\E_3^2 -\frac{\mathcal{C}^2 \, (\mathcal{C}-8 \,\p)}{4608}\,\E_4
-\frac{\mathcal{C} \, \left(3 \mathcal{C}^2+54 \,\s^2 (\mathcal{C}+6 \,\p)
-71 \mathcal{C} \,\p+186 \,\p^2
-360 \,\s^4\right)}{60480}\,\E_6. \notag
\end{align}

\medskip
\noindent
\underline{$n=7$}

\medskip
\noindent
\begin{align}
\label{5.8}
\T{7} =\, &-\frac{105}{8}  \mathcal{C}  \,\s\,\E_3\,\bm{u}^2
+\frac{105}{4} \mathcal{C} \,\,\p \,\s\,\E_3 \,\bm{u}' 
 \\  &  
+ \left(\frac{63}{8} \mathcal{C} \,\,\s (\mathcal{C}-4 \,\p)\,\E_3' 
-\frac{35}{16} \mathcal{C}^2  \,\s\,\E_2 \,\E_3
+\frac{35}{16} \mathcal{C}^2  \,\s\,\E_3
-\frac{105}{16} \mathcal{C}  \,\s \left(\mathcal{C} -8 \,\p+4 \,\s^2\right)\,\E_5\right)\,\bm{u}
+N_{7}\notag
\end{align}

where $N_7$ is given by: 

\begin{align}
\label{5.9}
N_7 = \, & -\frac{7}{16} \mathcal{C} \, \,\s (\mathcal{C}-7 \,\p) (\mathcal{C}-4 \,\p)\,\E_3''
-\frac{7}{32} \mathcal{C}^2 \,\s (\mathcal{C}-4 \,\p) \,\E_3'
+\frac{7}{8} \mathcal{C}  \,\s (\mathcal{C}-4 \,\p) \left(\mathcal{C}-8 \,\p+4 \,\s^2\right)\,\E_5'
\notag \\  &  
+\frac{35}{384}
   \mathcal{C}^2  \,\s (\mathcal{C}-3 \,\p)\,\E_2^2 \,\E_3
   -\frac{35}{384} \mathcal{C}^3  \,\s\,\E_2 \,\E_3+\frac{7}{384}
    \mathcal{C}^2  \,\s \left(\mathcal{C}+12 \left(\,\p-\,\s^2\right)\right)\,\E_3
\notag \\  &     
   -\frac{7}{384} \mathcal{C}^2  \,\s
   \left(\mathcal{C}-3 \left(\,\p+4 \,\s^2\right)\right)\,\E_3 \,\E_4
\notag \\  &        
   -\frac{7}{32} \mathcal{C}  \,\s \left(\mathcal{C}^2+8 \,\s^2
    (\mathcal{C}-8 \,\p)-12 \mathcal{C} \,\p+48 \,\p^2+16 \,\s^4\right)\,\E_7 .
\end{align}

\medskip
\noindent
\underline{$n=8$}

\medskip
\noindent

\begin{align}
\label{5.10}
\T{8} &= k_1 \, \bm{u}''' + k_{2} \,\E_2 \,\bm{u}'' + k_3 \bm{u} \, \bm{u}'' + 
k_4 \bm{u}'' + k_5 (\bm{u}')^2 + \notag\\
&\qquad k_{6} \,\E_2^2\, \bm{u}' + k_{7} \E_2 \bm{u}' 
+ k_{8} \,\E_4 \bm{u}' + k_{9} \bm{u}' \bm{u}^2 + k_{10} \,\E_2\, \bm{u} \bm{u}'+\notag\\
&\qquad k_{11} \bm{u} \bm{u}' + k_{12} \bm{u}' + k_{13} \bm{u}^4 
+ k_{14} \E_2 \bm{u}^3 + k_{15} \bm{u}^3 + k_{16} \E_2^2 \bm{u}^2 +  \notag\\
&\qquad k_{17} \E_2 \bm{u}^2 + k_{18} \E_4 \bm{u}^2 + k_{19} \bm{u}^2 
+ k_{20}\,\E_2^3 \bm{u} +k_{21} \E_2^2 \bm{u} + k_{22} \E_3^2 \bm{u} +  \notag\\
&\qquad k_{23} \E_2 \bm{u} + k_{24} \E_4 \bm{u}+ k_{25} \E_2 \E_4 \bm{u} 
+ k_{26} \E_6 \bm{u} + k_{27} \bm{u} + N_8,
\end{align}
with 
\begin{align}
&  k_1  = \, -\,\p^3 ,\qquad
  k_{2}  = \, \frac{\mathcal{C} \,\p^2}{2}  ,\qquad
  k_3  = \, 2 \,\p^2 ,\\
&  k_4  = \, -\frac{\mathcal{C} \,\p^2}{2},\qquad
  k_5  = \, \frac{3 \,\p^2}{2} ,\qquad
  k_{6}  = \, -\frac{1}{288} \mathcal{C} \,\p (7 \mathcal{C}+41 \,\p) ,\notag\\
&  k_{7}  = \, \frac{11 \mathcal{C}^2 \,\p}{96} ,\qquad
    k_{8}  = \, \frac{1}{288} \mathcal{C} \,\p \left(\mathcal{C}-43 \,\p+84 \,\s^2\right),\notag \\
&  k_{9}  = \, -\frac{3 \,\p}{2} ,\qquad
  k_{10}  = \, -\frac{3 \mathcal{C} \,\p}{4},\qquad
  k_{11}  = \, \frac{3 \mathcal{C} \,\p}{4} ,\notag\\
&  k_{12}  = \, -\frac{1}{96} \mathcal{C} \,\p \left(9 \mathcal{C}+28 \left(\,\s^2-\,\p\right)\right) ,\notag\\
&  k_{13}  = \, \frac{1}{8},\qquad
   k_{14}  = \, \frac{\mathcal{C}}{8},\qquad
   k_{15}  = \, -\frac{\mathcal{C}}{8} ,\notag\\
&  k_{16}  = \, \frac{1}{576} \mathcal{C} (7 \mathcal{C}+41 \,\p) ,\qquad
    k_{17}  = \, -\frac{11 \mathcal{C}^2}{192},\notag\\
&  k_{18}  = \, -\frac{1}{576} \mathcal{C} \left(\mathcal{C}-43 \,\p+84 \,\s^2\right),\notag \\
&  k_{19}  = \, \frac{1}{192} \mathcal{C} \left(9 \mathcal{C}+28 \left(\,\s^2-\,\p\right)\right) ,\notag\\
&  k_{20}  = \, \frac{\mathcal{C}
 \left(-3 \mathcal{C}^2+11 \mathcal{C} \,\p+76 \,\p^2\right)}{3456},\notag \\
&  k_{21}  = \, \frac{\mathcal{C}^2 (7 \mathcal{C}-106 \,\p)}{6912} ,\qquad
    k_{22}  = \, \frac{315 \mathcal{C}^2 \,\s^2}{2},\notag \\
&  k_{23}  = \, \frac{1}{576} \mathcal{C}^2 \left(3 \mathcal{C}+14 \left(\,\s^2-\,\p\right)\right) ,\notag\\
&  k_{24}  = \, -\frac{\mathcal{C}^2 \left(\mathcal{C}+62 \,\p-168 \,\s^2\right)}{6912} ,\notag\\
&  k_{25}  = \, \frac{\mathcal{C} \left(7 \mathcal{C}^2+\mathcal{C} \left(36 \,\s^2-19 \,\p\right)+228 \,\p \left(3 \,\p-8 \,\s^2\right)\right)}{17280} ,\notag\\
&  k_{26}  = \, \frac{\mathcal{C} \left(-\mathcal{C}^2+6
 \mathcal{C} \left(2 \,\p+7 \,\s^2\right)+4 \left(47 \,\p-30 \,\s^2\right)
  \left(\,\p-6 \,\s^2\right)\right)}{8640} ,\notag\\
&  k_{27}  = \, -\frac{\mathcal{C} \left(19 \mathcal{C}^2+\mathcal{C}
 \left(192 \,\s^2-156 \,\p\right)+288 \left(\,\p^2-3 \,\p \,\s^2+\,\s^4\right)\right)}{3456}.
\end{align}
The $N_{8}$ contribution is again a term that is independent on $a$. These explicit trace relations
have a uniform structure and are expected to admit  suitable generalizations for higher $n$.

\section{Generalization to $U(N)$ theories}
\la{sec:un}

The results we have presented so far have been computed for theories with $SU(2)$ gauge group.
It is interesting to  extend the  analysis to $U(N)$ theories in order to see whether new 
features arise. For completeness, we briefly discuss the special $U(1)$ case separately in 
App.~(\ref{app:u1}).

\subsection{The  pure gauge case}

\subsubsection{Undeformed ring: classical trace relations in $GL(N)$}

As is well known, the classical trace relations for a $N\times N$ matrix $\varphi$ are obtained from the trivial 
remark that the characteristic polynomial $\det(z-\varphi)$ has degree $N$. Thus, we expand the expression
\begin{align}
\la{6.1}
\det(z-\varphi) &= e^{\log\det(z-\varphi)} = e^{\tr\log(z-\varphi)} = z^{N}\,e^{\tr\log(1-\frac{\varphi}{z})} \\
&= z^{N}\,e^{\tr(-\frac{\varphi}{z}-\frac{\varphi^{2}}{2z^{2}}+\dots)} \notag \\
&= z^{N}\,\bigg[
1-\frac{t_1}{z}+\frac{\frac{t_1^2}{2}-\frac{t_2}{2}}{z^2}+\frac{-\frac{t_1^3}{6}+\frac
   {t_2 t_1}{2}-\frac{t_3}{3}}{z^3}+\frac{\frac{t_1^4}{24}-\frac{1}{4} t_2
   t_1^2+\frac{t_3
   t_1}{3}+\frac{t_2^2}{8}-\frac{t_4}{4}}{z^4}+\dots\bigg], \notag
\end{align}
where we have denoted $t_{n} = \tr\varphi^{n}$. The combinations multiplying negative powers of $z$ 
must be identically zero. In the $2\times 2$ case this means that the independent quantities are $t_{1}$
and $t_{2}$ and we must have 
\begin{align}
t_{3} &= \frac{3 \,t_1 \,t_2}{2}-\frac{\,t_1^3}{2}, &
t_{4} &= -\frac{\,t_1^4}{2}+\,t_2
   \,t_1^2+\frac{\,t_2^2}{2}, \notag \\
t_{5} &= \frac{5}{4} \,t_1 \,t_2^2-\frac{\,t_1^5}{4}, &
t_{6} &= -\frac{3}{4} \,t_2
   \,t_1^4+\frac{3}{2} \,t_2^2 \,t_1^2+\frac{\,t_2^3}{4}, \notag \\
t_{7} &= \frac{t_1^7}{8}-\frac{7}{8} \,t_2
   \,t_1^5+\frac{7}{8} \,t_2^2 \,t_1^3+\frac{7}{8} \,t_2^3 \,t_1, & 
t_{8} &= \frac{t_1^8}{8}-\frac{1}{2} \,t_2
   \,t_1^6-\frac{1}{4} \,t_2^2 \,t_1^4+\frac{3}{2} \,t_2^3 \,t_1^2+\frac{t_2^4}{8},
\end{align}
and so on. In the case of $U(3)$ we have three independent quantities  $t_{1}$,
$t_{2}$, and $t_{3}$ and, for instance, 
\begin{align}
t_{4} &=
\frac{t_1^4}{6}-\,t_2 \,t_1^2+\frac{4 \,t_3
   \,t_1}{3}+\frac{t_2^2}{2}, \notag \\
t_{5} &= \frac{t_1^5}{6}-\frac{5}{6} \,t_2 \,t_1^3+\frac{5}{6} \,t_3
   \,t_1^2+\frac{5 \,t_2 \,t_3}{6}, \notag \\
t_{6} &= \frac{t_1^6}{12}-\frac{1}{4} \,t_2 \,t_1^4+\frac{1}{3} \,t_3
   \,t_1^3-\frac{3}{4} \,t_2^2 \,t_1^2+\,t_2 \,t_3
   \,t_1+\frac{t_2^3}{4}+\frac{t_3^2}{3}. 
\end{align}

\subsubsection{One-instanton corrections in undeformed $SU(N)$  pure gauge theory}

In $SU(N)$ theory we have $t_{1}=0$ both at the classical { and} at the quantum level. The previous 
relations { at classical level} take the following simple form in the $SU(2)$ and $SU(3)$ cases
\be
\la{6.4}
SU(2): \qquad t_{3} = 0, \quad 
t_{4} = \frac{1}{2}\,t_2^2,  
\ee
and
\be
\la{6.5}
SU(3): \qquad 
t_{4} = \frac{t_2^2}{2}, \quad
t_{5} = \frac{5 \,t_2 \,t_3}{6}, \quad
t_{6} = \frac{t_2^3}{4}+\frac{t_3^2}{3}.
\ee
The quantum corrections to these relations should be captured by the relation generalizing (\ref{2.1}),
see for instance  \cite{Cachazo:2002ry},
\be
\vev{\tr{\frac{1}{z-\varphi}}} = \frac{P'(z)}{\sqrt{P^{2}(z)-4\Lambda^{2N}}}, \qquad
P(z) = z^{N}+\sum_{\ell=2}^N u_{\ell}\,z^{N-\ell}.
\ee
The first instanton correction appears in $\tr{\varphi^{2N}}$ and can be read from the large $z$
expansions
\begin{align}
\la{6.7}
SU(2): & \qquad 
\vev{\tr{\frac{1}{z-\varphi}}} = \frac{2}{z}-\frac{2 u_2}{z^3}+\frac{{4 \Lambda^{4}} +2 u_2^2}{z^5}+\dots,
\notag \\
SU(3): & \qquad 
\vev{\tr{\frac{1}{z-\varphi}}} = \frac{3}{z}-\frac{2 u_2}{z^3}-\frac{3 u_3}{z^4}
+\frac{2 u_2^2}{z^5}+\frac{5 u_2 u_3}{z^6}+\frac{{6 \Lambda^{6}} -2
   u_2^3+3 u_3^2}{z^7}+\dots.
\end{align}
Of course, the undeformed results in (\ref{6.7}) are in agreement with the  
general prediction in Eq.~(6.4) of \cite{Ashok:2016ewb}.

\subsubsection{All instanton relations in the $U(2)$ gauge theory on generic background}

The expressions in Sec.~(\ref{sec:empirical-su2}) admit a straightforward generalization to $U(2)$.
The only difference is the presence of the trivial trace 
\be
t_{1} = \vev{\tr\varphi} = a_{1}+a_{2}.
\ee
This expression is exact and does not receive instanton corrections. \footnote{This is clear from 
(\ref{B.9}) whose linear part in $z$ is only $\sum_{u=1}^{N}a_{u}$.} Thus, in terms of 
$t_{1}$ and $t_{2} = \T{2}$ we obtain after some trial and error the following list of exact relations
that we have checked at 10 instantons 
\begin{align}
\la{6.9}
\T{3} &= \frac{3 t_{1} t_{2}}{2}-\frac{t_{1}^3}{2}, \notag \\
\T{4} &= 4\, q-\frac{t_{1}^4}{2}+t_{1}^2
   t_{2}+\frac{t_{2}^2}{2}-\eps _1 \eps _2\,
   t_{2}', \notag \\
\T{5} &= 10\, q \left(t_{1}+\eps _1+\eps
   _2\right)-\frac{t_{1}^5}{4}+\frac{5 t_{1}
   t_{2}^2}{4}-\frac{5}{2}  \eps _1 \eps _2 t_{1}
   t_{2}' , \notag \\
\T{6} &= 12\, q\, t_{1}^2+30\, q\, t_{1} \left(\eps _1+\eps
   _2\right)+6\, q\, t_{2}+6\, q\, \left(3 \eps _1^2+4 \eps _2
   \eps _1+3 \eps _2^2\right)-\frac{3 t_{1}^4
   t_{2}}{4}+\frac{3 t_{1}^2 t_{2}^2}{2}\notag \\
   &-3\, t_{1}^2\,
   \eps _1 \eps _2\,
   t_{2}'+\frac{t_{2}^3}{4}+\eps _1^2 \eps _2^2\,
   t_{2}''-\frac{3}{2}  \eps _1 \eps _2\,t_{2}\,
   t_{2}', \notag \\
\T{7} &= 7\, q\, t_{1}^3+42\, q\, t_{1}^2
 \left(\eps _1+\eps_2\right)+21\, q\, t_{1} t_{2}+21\, q\, t_{1} \left(3
   \eps _1^2+4 \eps _2 \eps _1+3 \eps
   _2^2\right)+21\, q\, t_{2} \left(\eps _1+\eps
   _2\right)\notag \\
   &+7 q \left(4 \eps _1^3+7 \eps _2 \eps
   _1^2+7 \eps _2^2 \eps _1+4 \eps
   _2^3\right)+\frac{t_{1}^7}{8}-\frac{7 t_{1}^5
   t_{2}}{8}+\frac{7 t_{1}^3 t_{2}^2}{8}\notag \\
   &-\frac{7}{4}
    \eps _1 \eps _2 \,t_{1}^3 t_{2}'+\frac{7 t_{1}
   t_{2}^3}{8}+\frac{7}{2}  \eps _1^2 \eps
   _2^2 \,t_{1}\,t_{2}''-\frac{21}{4} \eps _1
   \eps _2 \,t_{1} t_{2} t_{2}', \notag \\
\T{8} &= 12\, q^2-2\, q t_{1}^4+28\, q\, t_{1}^3 \left(\eps _1+\eps
   _2\right)+36\, q\, t_{1}^2 t_{2}+4\, q\, t_{1}^2 \left(25
   \eps _1^2+33 \eps _2 \eps _1+25 \eps
   _2^2\right)\notag \\
   &+84\, q\, t_{1} t_{2} \left(\eps _1+\eps
   _2\right)+28\, q\, t_{1} \left(4 \eps _1^3+7 \eps _2
   \eps _1^2+7 \eps _2^2 \eps _1+4 \eps
   _2^3\right)+6\,q\, t_{2}^2-12\, q\, \eps _1 \eps _2
   t_{2}'\notag \\
   &+q\, t_{2} \left(52 \eps _1^2+72 \eps _2
   \eps _1+52 \eps _2^2\right)+8\, q\, \left(5 \eps _1^4+11
   \eps _2 \eps _1^3+15 \eps _2^2 \eps _1^2+11
   \eps _2^3 \eps _1+5 \eps
   _2^4\right)\notag \\
   &+\frac{t_{1}^8}{8}-\frac{t_{1}^6
   t_{2}}{2}-\frac{t_{1}^4 t_{2}^2}{4}+\frac{1}{2}
   \eps _1 \eps _2 \,t_{1}^4  t_{2}'+\frac{3
   t_{1}^2 t_{2}^3}{2}+6  \eps _1^2 \eps
   _2^2 \,t_{1}^2 t_{2}''-9  \eps _1 \eps _2\,t_{1}^2 t_{2}
   t_{2}'+\frac{t_{2}^4}{8}\notag \\
   & - \eps _1^3
   \eps _2^3\,t_{2}^{'''}+2  \eps _1^2 \eps _2^2\,t_{2}
   t_{2}''+\frac{3}{2} \eps _1^2 \eps _2^2
   \left(t_{2}'\right)^2-\frac{3}{2} \eps _1
   \eps _2 \, t_{2}^2 t_{2}', 
\end{align}
where, as usual, $X'\equiv q\partial_{q}X$. 

\subsubsection{$U(3)$ on a generic background: some surprise}

Repeating the same kind of analysis in the case of $U(3)$ we find something new. Now, the 
independent coordinates are 
\be
\la{6.10}
t_{1} = \vev{\tr\varphi} = a_{1}+a_{2}+a_{3}, \qquad
t_{2} = \T{2}, \qquad
t_{3} = \T{3}.
\ee
The relations we find for $\T{4}$ and $\T{5}$ are similar to the previous ones and read
\begin{align}
\la{6.11}
\T{4} &= \frac{t_{1}^4}{6}-t_{1}^2 t_{2}+\frac{4 t_{1}
   t_{3}}{3}+\frac{t_{2}^2}{2}-\eps _1 \eps _2\,
   t_{2}', \\
\T{5} &=\frac{t_{1}^5}{6}-\frac{5 t_{1}^3 t_{2}}{6}+\frac{5
   t_{1}^2 t_{3}}{6}+\frac{5 t_{2}
   t_{3}}{6}-\frac{5}{3} \eps _1 \eps _2\, t_{3}'. \notag
\end{align}
As a check of (\ref{6.11}), we can consider the undeformed limit $\bm{\eps}=0$ and 
restrict to $SU(3)$ setting $t_{1}=0$. This gives the classical relations (\ref{6.5}). This is correct
because, according to (\ref{6.7}), the first instanton correction is $\sim q = \Lambda^{6}$ and 
appears at dimension 6, {\em i.e.} in $\T{6}$.
Looking at $\T{6}$ for generic $\bm{\eps}$ one can try to mimic (\ref{6.11}) by fitting a generic
dimension 6 combination of monomials in $t_{1}$, $t_{2}$, $t_{3}$, and their derivatives. \footnote{
The derivative $q\partial_{q}$ increases effectively the dimension by 2 because such terms are always 
accompanied by explicit $\eps_{1}\eps_{2}$ factors. This reduces the set of monomials to be considered.
However, to be sure, we relaxed this hypothesis and checked it just at the end.
} Quite surprisingly, we could not find a solution in this way. Inspecting the low instanton corrections 
in full details, we noticed that the required missing structures appear in the following double trace
expectation values
\be
\la{6.12}
t_{n,m} = \vev{\tr\varphi^{n}\,\tr\varphi^{m}},
\ee
for the dimension 6 cases $(n,m) = (1,5), (2,4), (3,3)$. Usually, double traces like those in (\ref{6.12})
are not relevant because in the undeformed limit the supersymmetry algebra implies factorization of 
such correlators. However, with non zero $\eps_{1}$ and $\eps_{2}$, this is not the case. \footnote{
This peculiar  violation of factorization appears already in the NS limit, see for instance 
the discussion in  section 2.3 of \cite{Fucito:2005wc}, {\em e.g.} their Eq.(2.25).} Including 
$\vev{\tr\varphi\,\tr\varphi^{5}}$, $\vev{\tr\varphi^{2}\,\tr\varphi^{4}}$, and 
$\vev{(\tr\varphi^{3})^{2}}$ in the analysis, we find the following simple relation 
\be
\la{6.13}
\T{6} = 
-6 \, q+\frac{1}{12}\,t_{1}^{6}-\frac{1}{4}\,t_{1}^{4}\,t_{2}
+\frac{1}{3}\,t_{1}^{3}\,t_{3}
-\frac{3}{4}\,t_{1}^{2}\,t_{2}^{2}
+t_{1}\,t_{2}\,t_{3}
+\frac{1}{4}\,t_{2}^{3}+{\frac{1}{3}\,\bm{t_{3,3}}},
\ee
where we emphasized the double trace term. Besides, another outcome of the analysis is the vanishing of 
the following auxiliary combinations $K_{1}=K_{2}=0$,
%\be
%K_{1} = K_{2} = 0,
%\ee
where 
\begin{align}
\la{6.15}
K_{1} &=  -\frac{1}{5}\,t_{1}^6+t_{1}^4 t_{2}-t_{1}^3
   t_{3}-t_{1} t_{2} t_{3}+{\frac{6}{5}\,\bm{t_{1,5}}}, \notag \\
K_{2} &=  -\frac{1}{30}\,t_{1}^6-\frac{1}{6}\,t_{1}^3 t_{3}
+t_{1}^2   t_{2}^2-2\, \eps _1 \eps _2\, t_{1}^2\,   t_{2}'
-\frac{3}{2}\, t_{1} t_{2} t_{3}
+3 \,\eps _1 \eps _2\,t_{1}\, t_{3}'+\frac{1}{5}\,\bm{t_{1,5}}\notag \\
   &-\frac{1}{2}\,t_{2}^3-2 \,\eps_1^2\, \eps _2^2\, t_{2}''+3\, \eps _1 \eps_2 
   t_{2} \,t_{2}'+{\bm{t_{2,4}}}.
\end{align}
Again, we can look at (\ref{6.13}) in the undeformed $SU(3)$ limit. Using in this case 
$t_{1}=0$ and $\vev{(\tr\varphi^{3})^{2}}=\T{3}^{2}$, we recover the one-instanton correction
to the classical relation in (\ref{6.15}). Besides, the auxiliary relations (\ref{6.15}) are found to vanish
using again $t_{1,5}=t_{1}t_{5}=0$, $t_{2,4}=t_{2}t_{4}$ and replacing $t_{4}$ by means of 
(\ref{6.11}). One can also study the NS limit $\eps_{2}\to 0$ with the methods of 
App.~(\ref{app:pogho}) and again, one has full agreement exploiting the observation that the 
three dimension 6 double traces factorize in this limit, as we explicitly checked.

\subsection{Mass deformation: the $SU(3)$ $\mc N=2^{\star}$ theory on a generic background}

We can analyze in a similar way the $\mc N=2^{\star}$ theory. As an illustrative example
we take the gauge group to be $SU(3)$. Recalling the definitions in (\ref{zzz1}),
%To make the expressions more compact, it is convenient to introduce the quantities
%\be
%\la{6.16}
%\mc C = 4m^{2}-(\eps_{1}-\eps_{2})^{2},\qquad
%\text{p} = \eps_{1}\,\eps_{2},\qquad
%\text{s} = \eps_{1}+\eps_{2}.
%\ee
we have for $\T{4}$
\begin{align}
\la{6.17}
\T{4} &=
\frac{1}{2}\,t_{2}^{2}+\frac{1}{16}\,\mc C\, (\E_{2}-1)\,t_{2}
-\p\, t_{2}' \\
& +\frac{1}{256}\,\mc C^2 \E_{2}
-\frac{\mc C\,\E_{2}^2 (9\,\mc C-32\, \p)}{1536}
-\frac{\mc C\, \left(\mc C+32\, \left(\p-\s^2\right)\right)}{2560}
  +\frac{\mc C\,\E_{4}\, \left(9 \mc C-32\, \p-48\, \s^2\right)}{3840}.\notag
\end{align}
The expression (\ref{6.17}) is similar in structure to the analogous one for  gauge group 
$SU(2)$ in (\ref{5.3}). For $\T{5}$, we find 
\begin{align}
\la{6.18}
\T{5} &=\frac{5}{6}\, t_{2}\,t_{3}-\frac{5}{3}\, \p\,t_{3}'-\frac{5}{32}\, \mc C\,t_{3}
+\frac{5}{32}\, \mc C\, \E_{2}\, t_{3}
-\frac{45}{8} \,\mc C\, \s\, \E_{3}\,t_{2}-\frac{45}{128}\,\mc C^2\,\s\,\E_{3} \notag \\
&   +\frac{45}{128}\, \mc C^2\, \s\,\E_{2}\E_{3} 
    -\frac{15}{16} \,\mc C \, \s \left(\mc C-8\, \p+4\, \s^2\right)\,\E_{5}
    +\frac{15}{16}\, \mc C\, \s\, (3\,\mc  C-16\,\p)\,\E_{3}' ,
\end{align}
to be compared with (\ref{5.5}). Notice that there is no prefactor $\s$ in this case. This is because 
the trace is not trivial in the  undeformed theory. Finally, for $\T{6}$ we find 
\begin{align}
\la{6.19}
\T{6} &=\frac{1}{4}\,t_{2}^{3}+{\frac{1}{3}\,\bm{t_{3,3}}}
-\frac{3}{2}\, \p\,t_{2}\,t_{2}' 
+\p^2\,t_{2}'''+\frac{7}{64}\, \mc C\,(\E_{2}-1)\,t_{2}^{2} 
   -\frac{7}{32} \,\mc C\,\p\,(\E_{2}-1)\,t_{2}'  \notag \\
   &-\frac{3}{512}\,\mc C^2\,\E_{2}\,t_{2}
   -\frac{27}{2}\,\mc C\,\s\,\E_{3}\,t_{3}
   +\frac{1}{768}\, \mc C\, \left(3 \,\mc C-2 \left(\p+24\,\s^2\right)\right)\,\E_{4}\,t_{2} \notag \\
   &+\frac{\mc C\, (200\,\p-27\,\mc  C)}{3072}\, \E_{2}^{2}\,t_{2}
   +\frac{\mc C  \left(11 \mc C+64 \left(\s^2-\p\right)\right)}{1024}\,t_{2}\notag \\
   &+\frac{\mc C\, \left(653 \mc C^2-738\, \mc C \left(83\, \p-198\,
   \s^2\right)+19680\, \left(10\, \p^2-9\, \p\, \s^2-18\, \s^4\right)\right)}{1269760}\, \E_{2}\notag \\
   &+\frac{\mc C \, \left(-313\, \mc C^2+8\, \mc C\,
   \left(949\, \p-2079\, \s^2\right)-2240\, \left(10\, \p^2-9\, \p\, \s^2-18\, \s^4\right)\right)}{253952}\,
   \E_{4}\notag \\
   &+\frac{9\, \mc C \,
   \left(135\, \mc C^2+\mc C \,\left(5184\, \s^2-4980\, \p\right)+1600 
   \,\left(10\, \p^2-9\, \p\, \s^2-18\, \s^4\right)\right)}{3968}\,\E_{3}^{2}
  \notag \\
  &+\frac{\mc C
   \left(-99\, \mc C^2+\mc C\, \left(4303 \,\p-10014 \,\s^2\right)+16 \,\left(-878 \,
   \p^2+939 \,\p\, \s^2+1382\, \s^4\right)\right)}{126976}\notag \\
   &+\frac{\mc C\,
   \left(393\, \mc C^2+\mc C\, \left(5940 \,\s^2-3110\, \p\right)+800 \,\left(10\, \p^2-9 \,\p\, \s^2-18
   \,\s^4\right)\right)}{253952}\,\E_{2}^{2} \notag \\
   &+\frac{\mc C\, \left(531 \,\mc C^2-8\, \mc C\, \left(821 \,\p-3051\, \s^2\right)+320\, \left(52
   \,\p^2-363 \,\p \,\s^2+18 \,\s^4\right)\right)}{3809280}\, \E_{2}\,\E_{4}\notag \\
   &+\frac{75 \,\mc C \, \left(27\, \mc C^2+\mc C\, \left(2376 \,\s^2-996\, \p\right)+320\,
   \left(10\, \p^2-9\, \p \,\s^2-18\, \s^4\right)\right)}{15872}\,\E_{3}\notag \\
   &-\frac{3\, \mc C\,\left(27\, \mc C^2+\mc C\, \left(2376\,
   \s^2-996\, \p\right)+320\, \left(10 \,\p^2-9\, \p\, \s^2-18 \,\s^4\right)\right)}{15872}
   \, \E_{2}\,\E_{3} \notag \\
   &+\frac{9 \,\mc C \,\left(27\, \mc C^2+\mc C\,
   \left(2376\, \s^2-996\, \p\right)+320 \,\left(10\, \p^2-9 \,\p \,\s^2-18\, \s^4\right)\right)}{1984}
   \,\E_{3}' \notag \\
   &-\frac{\mc C\, \left(72\,
   \mc C^2+\mc C\, \left(599\, \p+198\, \s^2\right)-80 \,\left(38 \,\p^2+3\, \p\, \s^2
   +6 \,\s^4\right)\right)}{380928}
   \, \E_{2}^{3}, 
\end{align}
to be compared with (\ref{5.6}) in the $SU(2)$ case. Again, we have emphasized the double trace 
$t_{3,3}$ appearing in the first line of (\ref{6.19}).

\section{Trace relations in $\Omega$-deformed $\mc N=2^{\star}$ and  AGT correspondence}
\la{sec:agt}

In this section, we attempt to understand the origin of the trace relations in the 
$SU(2)$ $\mc N=2^\star$ theory on a generic $\Omega$-background
 exploiting AGT correspondence. To this aim, we begin with a brief review of the 
 analysis of the $N_{f}=4$ theory recently presented by 
Fucito, Morales and Poghossian (FMP) \cite{Fucito:2015ofa,Fucito:2016jng}.
Then, in a similar spirit, we discuss 
what happens in the case of the 
$\mc N=2^{\star}$ theory.

\subsection{Review of Fucito-Morales-Poghossian results} 

In the recent papers \cite{Fucito:2015ofa,Fucito:2016jng}, Fucito, Morales, and Poghossian 
discuss  the chiral correlators $\T{n}$ in the $SU(2)$ $\mc N=2$
gauge theory with $N_{f}=4$ fundamental hypermultiplets 
 in terms of four-point correlators in the 
Liouville theory with the insertion of the Liouville theory 
integrals of motion  introduced in \cite{Alba:2010qc}.
Here, we summarize their results to set up our notation and 
as a preliminary step for next application to 
the $\mc N=2^{\star}$ theory.

\subsubsection{CFT side}

The symmetry algebra of the Liouville theory is $\text{\sf Vir}\times \text{\sf Heis}$ with mode operators
\begin{align}
\la{7.1}
[L_{m}, L_{n}] &= (m-n)\,L_{m+n}+\frac{c}{12}\,m\,(m^{2}-1)\,\delta_{m+n,0}, \notag \\
[a_{m}, a_{n}] &= \frac{m}{2}\,\delta_{m+n,0}, \qquad [L_{m}, a_{n}]=0.
\end{align}
The standard AGT parametrization of the central charge is $c=1+6\,Q^{2}$ with $Q=b+b^{-1}$.
Primary fields are $V_{\alpha}=V_{\alpha}^{\text{\sf Vir}}\,V_{\alpha}^{\text{\sf Heis}}$ where 
$V_{\alpha}^{\text{\sf Vir}}$ is a primary with  conformal dimension $\Delta(\alpha)=
\alpha\,(Q-\alpha)$ and 
\be
\la{7.2}
V_{\alpha}^{\text{\sf Heis}}(z) = \exp\bigg(2\,i\,(\alpha-Q)\,\sum_{n<0}\frac{a_{n}}{n}\,z^{-n}
\bigg)\,\exp\bigg(2\,i\,\alpha\,\sum_{n>0}\frac{a_{n}}{n}\,z^{-n}\bigg).
\ee
%The Fock space is obtained by acting with $L_{n<0}$ and $a_{n<0}$ on the vacuum $|0\rangle$
%such that 
%\be
%\la{7.3}
%L_{m}|0\rangle = a_{n}|0\rangle=0,\qquad m\ge -1, \ n>0.
%\ee
As usual, primary states are defined by 
$
%\be
%\la{7.4}
|\alpha\rangle = V_{\alpha}(0)|0\rangle$
% and $, \qquad
%\langle\alpha| = \lim_{z\to\infty}z^{2\,\Delta(\alpha)}\,\langle 0 | V_{\alpha}(z),
%\ee
where $|0\rangle$ is the standard vacuum state.
We are interested in conformal blocks, {\em i.e.} four point functions with the exchange of
an $\alpha$-primary. It can be shown that 
\be
\la{7.5}
G(\alpha_{i}, \alpha|z) = \langle \alpha_{1}|V_{\alpha_{2}}(1)\,
V_{\alpha_{3}}(z) | \alpha_{4}\rangle_{\alpha} = (1-z)^{2\,\alpha_{2}\,(Q-\alpha_{3})}\,
G^{\text{\sf Vir}}(\alpha_{i}, \alpha | z),
\ee
where $G^{\text{\sf Vir}}(\alpha_{i}, \alpha | z)$ is the standard conformal block.
Following
FMP, we can  introduce Liouville integrals of motion according to, {\em cf.} \cite{Alba:2010qc},
\begin{align}
\la{7.6}
I_{2} &= L_{0}-\frac{c}{24}+2\,\sum_{k=1}^{\infty}a_{-k}\,a_{k}, \notag \\
I_{3} &= \sum_{k\in\mathbb Z\backslash \{0\}}
a_{-k}\,L_{k}+2\,i\,Q\,\sum_{k=1}^{\infty}
k\,a_{-k}\,a_{k}+\frac{1}{3}\sum_{i+j+k=0}a_{i}\,a_{j}\,a_{k}, \notag\\
I_{4} &= 2\,\sum_{k=1}^{\infty}L_{-k}\,L_{k}+L_{0}^{2}-\frac{c+2}{12}\,L_{0}+\text{other
terms involving \sf Heis}.
\end{align}
These may be inserted in the conformal block to build the new quantities
\be
\la{7.7}
G_{n}(\alpha_{i}, \alpha|z) =  \langle \alpha_{1}|V_{\alpha_{2}}(1)\,I_{n}\,
V_{\alpha_{3}}(z) | \alpha_{4}\rangle_{\alpha}.
\ee
Exploiting the $\text{\sf Vir}\times \text{\sf Heis}$ algebra it is possible to prove relations like 
\be
\la{7.8}
G_{2} = \bigg(z\,\partial_{z}+\Delta_{3}+\Delta_{4}-\frac{c}{24}\bigg)\,G.
\ee
and similar (more involved) ones for higher  $G_{n}$'s. 

\subsubsection{Chiral traces and AGT correspondence}

 The AGT correspondence relates the four-point conformal block of the Liouville theory to the partition function of the $\mc N = 2$ supersymmetric $SU(2)$ gauge theory with four fundamentals. The
 parameter $q$  is identified with the harmonic ratio $z$ parametrizing the positions of vertex insertions.
The four dimensions $\Delta_{i}$ are functions
of the masses while the internal dimension $\Delta$ is function of the vacuum expectation value $a$.
Finally, the deformation parameters appear in the central charge formula according to 
the proportionality $\eps_{1}:\eps_{2} = b:b^{-1}$ \cite{Alday:2009aq}. This dictionary allows to 
rewrite relations like (\ref{7.8}) in terms of gauge theory parameters {once a precise correspondence between $\T{n}$ and $G_{n}$ is established.} The relations proposed by FMP 
are
\begin{align}
\la{7.9}
\T{2} &\equiv \bm{u} =  -2\,\frac{G_{2}}{G}-\frac{1}{12}, \qquad
\T{3} = 6\,i\,\frac{G_{3}}{G}, \notag \\
\T{4} &= 2\,\p^{2}\,\frac{G_{4}}{G}-\frac{\p}{4}\,\bm{u}+\frac{\s^{2}\,(\p+\s^{2})}{8}.
\end{align}
With this identifications, it is possible to use relations like (\ref{7.8}) and write $\T{n}$ in terms
of  multiple $q\partial_{q}$ derivatives of $\log Z$ where $Z$ is the partition function of the gauge theory.
In particular, Matone's relation \cite{Matone:1995rx} reads in 
this context
\be
\la{7.10}
\T{2} = -2\,\p\,q\,\partial_{q}\log Z.
\ee
We remark that the precise form of $I_{n}$ {is not enough} to predict $\T{n}$ because we need the 
map from the quantities $G_{n}$ to 
the generators of the chiral ring, as in (\ref{7.9}). 
Nevertheless, it is important to emphasize that 
 the presence of higher powers of $L_{0}$ immediately 
implies the occurrence of further $q\partial_{q}$-derivatives in higher traces. 

\subsection{The $\mc N=2^{\star}$ theory}

\subsubsection{Hints from Matone's relation}

Let us analyze the $\mc N=2^{\star}$ theory on a generic $\Omega$-deformation
starting from the generalized Matone's relation  \cite{Matone:1995rx,Flume:2004rp}
\be
\la{7.11}
\T{2} = 2\,a^{2}-2\,\eps_{1}\,\eps_{2}\,\,q\,\partial_{q} \log Z_{\text{\sf inst}},
\ee
where $Z_{\rm inst}$ is the instanton partition function of the gauge theory. The AGT dictionary
reads
\begin{align}
\la{7.12}
& c=1+6\,Q^{2}, \qquad Q = b+b^{-1},\qquad b = \sqrt{\eps_{2}/\eps_{1}},\\
\la{7.13}
& \Delta = \frac{Q^{2}}{4}-\frac{a^{2}}{\eps_{1}\,\eps_{2}},\qquad 
\Delta_{m} = \frac{Q^{2}}{4}-\frac{m^{2}}{\eps_{1}\,\eps_{2}}. 
\end{align}
The instanton partition function may be computed in terms of the {\sf Vir} algebra with central charge $c$
according to \footnote{The Dedekind $\eta$ function is $\eta=q^{\frac{1}{24}}\prod_{k=1}^{\infty}
(1-q^{k})$. Notice that  $q\frac{d}{dq}\log\prod_{k=1}^{\infty}(1-q^{k}) = 
\frac{1}{24}\,(\E_{2}-1)$.
}
\be
\la{7.14}
Z_{\text{\sf inst}} = \left[q^{-\frac{1}{24}}\,\eta(\tau)\right]^{2\,\Delta_{m}-1}\,\mc F^{\Delta}_{\Delta_{m}}(q).
\ee
Here,  the so-called torus one-point functions $\mc F^{\Delta}_{\Delta_{m}}(q)$ is 
\be
\la{7.15}
\mc F^{\Delta}_{\Delta_{m}}(q) = \tr_{\Delta}\bigg(\mc O_{\Delta_{m}}\,q^{L_{0}-\Delta}\bigg),
\ee
where the trace is over all descendants of the Virasoro primary 
$\mc O_{\Delta}$. \footnote{
To check normalization, it is useful to recall 
the following expansions at small $q$ or large intermediate dimension $\Delta$
\be
\notag
\mc F^{\Delta}_{\Delta_{m}}(q) = 1+\mc O(q), \qquad 
\mc F^{\Delta}_{\Delta_{m}}(q) = \frac{q^{\frac{1}{24}}}{\eta(\tau)}\bigg[1+\mc O\left(
\frac{1}{\Delta}\right)\bigg].
\ee
}
After these preliminary definitions we can plug (\ref{7.14}) into (\ref{7.11}) to find \footnote{
We can check  (\ref{7.16}) in the large $\Delta$ limit or -- what is the same -- at large $a$ 
in the gauge theory. Using the second expansion in (\ref{7.16}) we have 
\be
\T{2} = 2\,a^{2}-2\,\eps_{1}\,\eps_{2}\bigg[(2\Delta_{m}-2)\,\frac{\E_{2}-1}{24}+\mc O(a^{-1})\bigg] 
= 2\,a^{2}+(4\,m^{2}-(\eps_{1}-\eps_{2})^{2})\,\frac{\E_{2}-1}{24}+\mc O(a^{-1}).\notag
\ee
This is in agreement with the undeformed limit in (\ref{2.10}) and also with explicit similar expressions
with non zero $\bm{\eps}$.
}
\begin{align}
\la{7.16}
\T{2} &= 2\,a^{2}-2\,\eps_{1}\,\eps_{2}\bigg[(2\Delta_{m}-1)\,\frac{\E_{2}-1}{24}+
q\,\partial_{q}\log \mc F^{\Delta}_{\Delta_{m}}(q)\bigg] 	\notag \\
%\end{align}
%Eq. (\ref{7.16}) may be written as follows
%\begin{align}
%\la{8.18}
%\T{2} &= 
%%2\,a^{2}+(4\,m^{2}-\eps_{1}^{2}-\eps_{2}^{2})\,\frac{\E_{2}-1}{24}
%-2\,\eps_{1}\,\eps_{2}\,q\,\partial_{q}\log \mc F^{\Delta}_{\Delta_{m}}(q) \notag \\
%&= 2\,a^{2}+(4\,m^{2}-\eps_{1}^{2}-\eps_{2}^{2})\,\frac{\E_{2}-1}{24}
%-2\,\eps_{1}\,\eps_{2}\,q\,\partial_{q}\log\bigg[
%q^{-\Delta+\frac{c}{24}}
%\tr_{\Delta}\bigg(\mc O_{\Delta_{m}}\,q^{L_{0}-\frac{c}{24}}\bigg)\bigg]\notag \\
%&=
&= (4\,m^{2}-\eps_{1}^{2}-\eps_{2}^{2})\,\frac{\E_{2}-1}{24}-\frac{\eps_{1}\,\eps_{2}}{12}
-2\,\eps_{1}\,\eps_{2}\frac{\tr_{\Delta}\bigg(\mc O_{\Delta_{m}}\,
(L_{0}-\frac{c}{24})\,q^{L_{0}-\frac{c}{24}}\bigg)}
{\tr_{\Delta}\bigg(\mc O_{\Delta_{m}}\,q^{L_{0}-\frac{c}{24}}\bigg)}. 
\end{align}
Eq. (\ref{7.16}) may be considered as  the $\mc N=2^{\star}$ version of the first relation in (\ref{7.9}). Here 
the role of $G$ and $G_{2}$ is played by the one-point torus function with possible insertion 
of the Virasoro part of $I_{2}$ in (\ref{7.6}). In other words, (\ref{7.16}) is totally similar to (\ref{7.9})
when written in the form 
\be
\la{7.17}
\T{2} = 
-2\,\eps_{1}\,\eps_{2}\frac{G_{2}^{\star}}{G^{\star}}+
(4\,m^{2}-\eps_{1}^{2}-\eps_{2}^{2})\,\frac{\E_{2}-1}{24}-\frac{\eps_{1}\,\eps_{2}}{12},
\ee
where, see (\ref{7.5}) and (\ref{7.7}), 
\be
\la{7.18}
G^{\star} = \tr_{\Delta}\bigg(\mc O_{\Delta_{m}}\,q^{L_{0}-\frac{c}{24}}\bigg),
\qquad
G^{\star}_{2} = \tr_{\Delta}\bigg(\mc O_{\Delta_{m}}\,I_{2}^{\text{\sf Vir}}
\,q^{L_{0}-\frac{c}{24}}\bigg)
= q\,\partial_{q}\,G^{\star}.
\ee

\subsubsection{A consistency check: prediction of the leading terms in $\T{n}$}

The trace relations for $\T{n}$ have been fully written out in Sec.~(\ref{sec:empirical}) up to $n=8$. 
Special {\em leading} terms are dimension $n$ monomials according to the 
weights
$[\bm{u}]=2$ and $[q\,\partial_{q}]=2$. 
This definition isolates simple non trivial parts of $\T{n}$ that do not
involve Eisenstein sums and are non zero  for even $n$. The first cases 
are (dots stand for lower dimension terms)
\begin{align}
\la{7.19}
\T{2} &= \bm{u},\qquad (\text{by definition}), \notag \\
\T{4} &= -\p\,\bm{u'}+\frac{1}{2}\,\bm{u}^{2}+\dots, \notag\\
\T{6} &= \p^{2}\,\bm{u''}-\frac{3}{2}\,\p\,\bm{u\,u'}+\frac{1}{4}\,\bm{u}^{3}+\dots, \notag \\
\T{8} &= -\p^{3}\,\bm{u'''}+2\,\p^{2}\,\bm{u\,u''}+\frac{3}{2}\,\p^{2}\,\bm{(u')^{2}}
-\frac{3}{2}\,\p\,\bm{u^{2}\,u'}+\frac{1}{8}\,\bm{u}^{4}+\dots. 
\end{align}
It is natural to expect that these terms come from
the genuine $n$-th order part of the integral of motion $I_{n}$, {\em i.e.} the part that cannot be modified with a mixing from 
lower order integrals of motion. In $\T{2n}$ this is nothing but the operator $L_{0}^{n}$. Thus, a natural
conjecture is 
\be
\la{7.20}
\T{2\,n}_{\text{\sf leading}} = 2\,\frac{(-\p\,q\,\partial_{q})^{n}\,G^{\star}_{\text{\sf leading}}}
{G^{\star}_{\text{\sf leading}}}, \qquad G^{\star}_{\text{\sf leading}} = \exp\bigg(
-\frac{U}{2\,\p}\bigg), \quad q\,\partial_{q} U = \bm{u}.
\ee
In fact, we can check that (\ref{7.20})
 works perfectly. For instance, 
\begin{align}
\la{7.21}
\T{4} &= 2\,\p^{2}\,e^{-\frac{U}{2\,\p}}\,\bigg[e^{\frac{U}{2\,\p}}\bigg]'' = -\p\,U''+
\frac{1}{2}(U')^{2} = -\p\,\bm{u'}+\frac{1}{2}\,\bm{u}^{2}, \notag \\
\T{6} &= -2\,\p^{3}\,e^{-\frac{U}{2\,\p}}\,\bigg[e^{\frac{U}{2\,\p}}\bigg]''' = 
\p^{2}\,U'''-\frac{3}{2}\,\p\,U'\,U''+\frac{1}{4}(U')^{3} \notag \\
&= \p^{2}\,\bm{u''}-\frac{3}{2}\,\p\,\bm{u\,u'}+\frac{1}{4}\,\bm{u}^{3},
\end{align}
and so on. Notice that  (\ref{7.20}) is a non trivial constraint as further 
discussed in App.~(\ref{app:nontrivial}).

\subsubsection*{Structure of subleading terms}

The subleading terms in the trace relations depend on the precise mapping between the integrals of motion
and the generators of the chiral ring. Nevertheless, they are captured by suitable insertions of 
powers of $L_{0}$. We can consider for instance the  list of 27 $\bm{u}$-dependent 
contributions to $\T{8}$ in (\ref{5.10}). Replacing $\bm{u}$ by $(\log Z_{\text{\sf inst}})'$ using
(\ref{7.11}) we see that $\T{8}$ is a linear combination of terms $\sim Z_{\text{\sf inst}}^{(k)}/
Z_{\text{\sf inst}}$ that encode all the nonlinearities in $\bm{u}$ and its derivatives. 
Up to $a$ dependent mixing terms we have for instance
\begin{align}
& \T{8} = 2\,\p^{4}\,\frac{Z_{\text{\sf inst}}^{(4)}}{Z_{\text{\sf inst}}}-\mc C\,\p^{3}\,(\E_{2}-1)\,
\frac{Z_{\text{\sf inst}}^{(3)}}{Z_{\text{\sf inst}}}+
(c_{1}+c_{2}\,\E_{2}+c_{3}\,\E_{2}^{2}+c_{4}\,\E_{4})\,\p^{2}
\,\frac{Z_{\text{\sf inst}}^{(2)}}{Z_{\text{\sf inst}}}\notag \\
& \ \ \ +
(c_{5}+c_{6}\,E_{2}+c_{7}\,\E_{2}^{2}+c_{8}\,\E_{4}+c_{9}\,\E_{2}^{3}
+c_{10}\,\E_{2}\,\E_{4}+c_{11}\,\E_{3}^{2}+c_{12}\,\E_{6})\,
\p\,\frac{Z_{\text{\sf inst}}^{(1)}}{Z_{\text{\sf inst}}}+\dots,
\end{align}
where $c_{i}$ may be expressed as linear combinations of the $k_{i}$ in (\ref{5.10}).

\subsubsection*{Universality of the leading terms}

A further consistency check of the AGT interpretation of trace relations
follows from the following argument. Let us consider AGT 
correspondence for pure gauge $SU(2)$ theory. The 
relevant CFT object is the irregular conformal block \cite{Gaiotto:2009ma}. 
 To define it, one starts with  the Whittaker vector
\be
\la{7.23}
|\Delta, \Lambda^{2}\rangle = v_{0}+\Lambda^{2}\,v_{1}+\Lambda^{4}\,v_{2}+\dots,
\ee
where $v_{0}\equiv |\Delta\rangle$ is a Virasoro highest weight state and the components 
$v_{n}$ are determined by the conditions
\be
\la{7.24}
L_{1}\,v_{n} = v_{n-1},\qquad L_{2}\,v_{n}=0.
\ee
The instanton partition function of the pure gauge $SU(2)$ theory is then simply
\be
\la{7.25}
Z_{\text{\sf inst}} = \langle \Delta, \Lambda^{2}\,|\,\Delta, \Lambda^{2}\rangle = 
\sum_{n=0}^{\infty} \Lambda^{4\,n} \lVert v_{n} \rVert^{2},
\ee
where $\Lambda^{4}$ is identified with the instanton counting parameter. Of course, the 
Virasoro data $\Delta$ and $c$ are translated in gauge theory parameters $a, \eps_{1}, \eps_{2}$ with the 
usual AGT dictionary, see (\ref{7.12}) and (\ref{7.13}).
If we now assume that the chiral observables are obtained by insertion of integrals of motion,
we have  the following schematic relation for the leading terms
\be
\T{2n}_{\text{\sf leading}} \sim \frac{\langle \Delta, \Lambda^{2}\,|L_{0}^{n}|\,\Delta, \Lambda^{2}\rangle}
{\langle \Delta, \Lambda^{2}\,|\,\Delta, \Lambda^{2}\rangle}\sim
\frac{(q\,\partial_{q})^{n} Z_{\text{\sf inst}}}{Z_{\text{\sf inst}}}.
\ee
This is same as in $\mc N=2^{\star}$ and leads to the conclusion 
\be
\left. \T{2n}_{\text{\sf leading}}\right|_{\text{pure gauge}} = 
\left. \T{2n}_{\text{\sf leading}}\right|_{\mc N=2^{\star}}.
\ee
From inspection of the explicit leading terms in (\ref{3.3}) and  comparing with 
(\ref{7.19}), we confirm that this is indeed true in our localization computation.

\section*{Acknowledgments}

M.B. is grateful to  A. Lerda for suggestions and clarifications. We thank
M. Bill\`o,  M. Frau, and M. Matone for  useful discussions.

\appendix

\section{Eisenstein series}
\la{app:eisen}

The Eisenstein series $\E_{2}$, $\E_{4}$, and $\E_{6}$ 
\cite{koblitz2012introduction} admit the  representation
\begin{align}
& \E_2(q) = 1 - 24 \sum_{n=1}^{\infty} \sigma_1(n)\, q^n 	& =\, &
 1-24\, q-72 \,q^2-96 \,q^3-168 \,q^4+\mc O(q^5)\notag\\
& \E_4(q) = 1 + 240 \sum_{n=1}^{\infty} \sigma_3(n)\, q^n 	& =\,&
 1+240\, q+2160 \,q^2+6720 \,q^3+17520 \,q^4+\mc O(q^5)\notag\\  
& \E_6(q) = 1 - 504 \sum_{n=1}^{\infty} \sigma_5(n)\, q^n 	& =\, &
 1-504 \,q-16632 \,q^2-122976 \,q^3-532728 \,q^4+\mc O(q^5),
\end{align}
where $\sigma_k(n)$ is the divisor function
$\sigma_{k}(n) = \sum_{d|n}d^{k}$. The Eisenstein series have well defined properties
under the modular group $SL(2, \mathbb{Z})$. In particular, $\E_{4}$ and $\E_{6}$
are modular forms of weight 4 and 6, while $\E_{2}$ is a quasi modular form of 
degree 2. 
For the purposes of this paper, we shall also introduce the following non-standard objects.
\be
\E_{2k+1} \stackrel{\rm def}{=}\sum_{n=1}^{\infty} \sigma_{2k}(n)\,q^{n},
\ee
or explicitly
\begin{align}
& \E_3(q) = \sum_{n=1}^{\infty} \sigma_2(n)\, q^n 	= 
 q+5 \,q^2+10 \,q^3+21 \,q^4+\mc O(q^5)\notag\\
& \E_5(q) = \sum_{n=1}^{\infty} \sigma_4(n)\, q^n 	=  
q+17 \,q^2+82 \,q^3+273 \,q^4+\mc O(q^5).
\end{align}
These series are not natural from the point of view of modular transformations, but are somewhat
expected in the spirit of the derivation discussed in Sec. \ref{sec:undef}, see
footnote \ref{f6}.

\section{Chiral observables from localization}
\la{app:localization}

The deformed partition function as well the chiral traces 
may be computed systematically by localization, see for instance \cite{Billo:2012st} 
and references therein. Here, we briefly discuss 
the  illustrative case of the 
$\mc N=2^{*}$ gauge theory with gauge group $U(N)$. Other simpler cases may be 
treated in quite similar way. Focusing on the algorithmic implementation the $k$-instanton corrections
to the partition function $Z = 1+\sum_{k=1}^{\infty} Z_{k}\,q^{k}$ are
 obtained as 
\be
\la{B.1}
Z_{k}(\bm{a}, m, \eps_{1}, \eps_{2}) = \sum_{|Y_{1}|+\cdots+|Y_{N}|=k} 
Z(Y_{1}, \dots, Y_{N}),
\ee
where $\bm{a}=\vev{\varphi}=(a_{1}, \dots, a_{N})$ and
we sum over all $N$-tuples  $(Y_{1}, Y_{2}, \dots, Y_{N})$ of Young tableaux with a total of $k$ blocks 
($|Y|$ is the number of blocks in a tableau). For each  $N$-tuple
$(Y_{1}, \dots, Y_{N})$, we build the symbol
\be
\la{B.2}
V = T_{a_{1}}\,V(Y_{1})+\cdots +T_{a_{N}}\,V(Y_{N}),
\ee
where
\be
\la{B.3}
V(Y) = \sum_{ij} T_{\eps_{1}}^{i-1}\, T_{\eps_{2}}^{j-1}.
\ee
In these expressions the symbols $T_{x}$ must be thought as Abelian {characters}  with 
\be
\la{B.4}
T_{x}\,T_{y} = T_{x+y}, \qquad T_{x}^{n} = T_{nx}, \qquad T_{x}^{*} = T_{-x}, \qquad 
\text{and so on.}
\ee
The sum over $i,j$ in (\ref{B.3}) is over the blocks of $Y$,  $i\ge 1$ is the row and $j\ge 1$ is the column.
After computing $V$, we introduce the universal 
object $W = T_{a_{1}}+\cdots+T_{a_{N}}$, 
and evaluate
\begin{align}
\la{B.5}
T_{\text{gauge}} &= -VV^{*}(1-T_{\eps_{1}})(1-T_{\eps_{2}})+V^{*}W+VW^{*}\, T_{\eps_{1}}T_{\eps_{2}}, \notag \\
T_{\text{matter}} &= T_{m-\frac{\eps_{1}+\eps_{2}}{2}}\, T_{\text{gauge}}.
\end{align}
These expressions may be obtained from the exact sequence associated with an instanton 
\cite{Flume:2004rp,Bruzzo:2002xf,Shadchin:2004yx}.
The results may be written in the form 
\be
\la{B.6}
T_{\text{gauge}} = \sum_{i}\,n_{i}\,T_{x_{i}}, \qquad 
T_{\text{matter}} = \sum_{i}\,m_{i}\,T_{y_{i}}, \qquad 
n_{i}, m_{i}\in\mathbb Z,
\ee
From (\ref{B.6}), we can read the partition function associated with $N$-tuple of  Young tableaux
as 
\be
Z(Y_{1}, \dots, Y_{N}) = \frac{\prod_{i}y_{i}^{m_{i}}}{\prod_{i}x_{i}^{n_{i}}}.
\ee
%As we have explained, the partition function is schematically
%\be
%\la{F.9}
%Z = 1+\sum_{k=1}^{\infty} Z_{k}\,q^{k},\qquad 
%Z_{k} = \sum_{|\bm{Y}|=k} \mc Z(\bm{Y})
%\ee
%where $\bm{Y}=(Y_{1}, \dots, Y_{N})$ and $|\bm{Y}|=\sum_{u=1}^{N}|Y_{u}|$. 
For an observable
$\mc O$ we shall introduce a specific function $\mc O(\bm{Y})$ and evaluate 
\be
\la{B.8}
\vev{\mc O} = \frac{1}{Z}\,\sum_{k=0}^{\infty}q^{k}\sum_{|\bm{Y}|=k} \mc Z(\bm{Y})\,
\mc O(\bm{Y}).
\ee
In the case of $\tr e^{z\,\varphi}$ the recipe is to use the following $\mc O(\bm Y)$,
see for instance \cite{Losev:2003py},
\be
\la{B.9}
\mc O(\bm{Y}) = \sum_{u=1}^{N}\bigg[
e^{z\,a_{u}}+(1-e^{z\,\eps_{1}})(1-e^{z\,\eps_{2}})\,
\sum_{i_{\rm R}=1}^{\#\, \text{rows of}\, Y_{u}}\sum_{i_{\rm C}=1}^{\#\,
\text{cols of}\, Y_{u}}
e^{z\,(a_{u}+(i_{\rm R}-1)\eps_{1}+(i_{\rm C}-1)\eps_{2})}\bigg].
\ee
Expanding in powers of $z$ we compute $\T{n}$. Multiple traces are evaluated in the same
way by considering, {\em e.g.},  the map 
$\vev{\tr\varphi^{n}\,\tr\varphi^{m}}\longrightarrow 
\mc O_{n}(\bm{Y})\mc O_{m}(\bm{Y})$, and so on.

\section{Higher order traces in deformed $SU(2)$ $\mc N=2$ theory}
\la{app:pure}

We list here some long expansions for higher traces in the deformed $SU(2)$ $\mc N=2$ theory.
These are organized in a double expansion at large $a$ around the undeformed point 
$\eps_{1}=\eps_{2}=0$.
\begin{align}
\T{4} =\, & 
2 \,a^4+6 q+\frac{9 \,q^2}{8 \,a^4}+\frac{7 \,q^3}{8 \,a^8}
+\frac{2145 \,q^4}{2048 \,a^{12}} + \dots 
\notag \\ \notag & +
\eps_1 \eps_2 \left(-\frac{q}{\,a^2}-\frac{7 \,q^2}{8 \,a^6}
-\frac{63 \,q^3}{32 \,a^{10}}-\frac{5315 \,q^4}{1024 \,a^{14}} + \dots\right)
\\ \notag & 
+(\eps_1+\eps_2)^2 \left(\frac{q}{2 \,a^2}+\frac{25 \,q^2}{16\,a^6}
+\frac{267 \,q^3}{64 \,a^{10}}+\frac{22529 \,q^4}{2048 \,a^{14}} + \dots\right)
\\ \notag & +
(\eps_1 \eps_2)^2 \left(\frac{43 \,q^2}{128 \,a^8}+\frac{373 \,q^3}{128 \,a^{12}}
+\frac{288189 \,q^4}{16384 \,a^{16}} + \dots\right)
\\ \notag & +
\eps_1 \eps_2 (\eps_1+\eps_2)^2 \left(-\frac{q}{4 \,a^4}-\frac{77 \,q^2}{32\,a^8}
-\frac{33 \,q^3}{2 \,a^{12}}-\frac{367069 \,q^4}{4096 \,a^{16}} + \dots\right)
\\ \notag & +
(\eps_1 \eps_2)^3 \left(-\frac{29 \,q^2}{256 \,a^{10}}
-\frac{1939 \,q^3}{512 \,a^{14}}-\frac{1699071 \,q^4}{32768 \,a^{18}} + \dots\right)
\\  & +
(\eps_1 \eps_2)^2 (\eps_1+\eps_2)^2 \left(\frac{885 \,q^2}{512 \,a^{10}}
+\frac{40503 \,q^3}{1024 \,a^{14}}+\frac{30341215 \,q^4}{65536 \,a^{18}} + \dots\right)+ \dots
\end{align}

%%%%%%%%%%%%%%%%%%%%%%%%%%%%%%%%%%%%%%%%%%%%%%%%%%%%%%%%%%%%%%%%%%%%%%%%%%%%

\begin{align}
\T{6} =\, &  
2 \,a^6+15 \,a^2 q+\frac{135 \,q^2}{16 \,a^2}+\frac{125 \,q^3}{32 \,a^6}
+\frac{16335 \,q^4}{4096 \,a^{10}} + \dots
\notag \\ \notag & +
\eps_1 \eps_2 \left(-15 q-\frac{15 \,q^2}{4 \,a^4}-\frac{105 \,q^3}{16 \,a^8}
-\frac{2025 \,q^4}{128 \,a^{12}} + \dots\right)
\\ \notag & +
(\eps_1+\eps_2)^2 \left(\frac{75 q}{4}+\frac{135 \,q^2}{32 \,a^4}
+\frac{735 \,q^3}{64 \,a^8}+\frac{124575 \,q^4}{4096 \,a^{12}} + \dots\right)
\\ \notag & +   
(\eps_1 \eps_2)^2 \left(\frac{q}{\,a^2}+\frac{545 \,q^2}{256 \,a^6}
+\frac{5139 \,q^3}{512 \,a^{10}}+\frac{1645387 \,q^4}{32768 \,a^{14}} + \dots\right) 
\\ \notag & +
\eps_1 \eps_2 (\eps_1+\eps_2)^2 \left(-\frac{3 q}{4 \,a^2}
-\frac{405 \,q^2}{64 \,a^6}-\frac{639 \,q^3}{16 \,a^{10}}-\frac{1759671 \,q^4}{8192 \,a^{14}} + \dots\right)
\\ \notag & +
(\eps_1 \eps_2)^3 \left(-\frac{409 \,q^2}{512 \,a^8}
-\frac{13887 \,q^3}{1024 \,a^{12}}-\frac{9539613 \,q^4}{65536 \,a^{16}} + \dots\right) 
\\  & +
 (\eps_1 \eps_2)^2 (\eps_1+\eps_2)^2 \left(\frac{q}{4 \,a^4}
 +\frac{7023 \,q^2}{1024 \,a^8}+\frac{213033 \,q^3}{2048 \,a^{12}}+\frac{142336817 \,q^4}{131072 \,a^{16}} + \dots\right)  + \dots
\end{align}

%%%%%%%%%%%%%%%%%%%%%%%%%%%%%%%%%%%%%%%%%%%%%%%%%%%%%%%%%%%%%%%%%%%%%%%%%%%%

\begin{align}
\T{7} =\, &  
(\eps_1+\eps_2) \left(42 \,a^2 q+\frac{21 \,q^2}{\,a^2}
+\frac{105 \,q^3}{16 \,a^6}+\frac{189 \,q^4}{32 \,a^{10}} + \dots\right)
\notag \\ \notag & +
\eps_1 \eps_2 (\eps_1+\eps_2) \left(-35 q-\frac{21 \,q^3}{8 \,a^8}
-\frac{21 \,q^4}{2 \,a^{12}} + \dots\right) 
\\ \notag & +
(\eps_1+\eps_2)^3 \left(28 q+\frac{21 \,q^2}{4 \,a^4}
+\frac{441\,q^3}{32 \,a^8}+\frac{1155 \,q^4}{32 \,a^{12}} + \dots\right)
\\ \notag & +
(\eps_1 \eps_2)^2 (\eps_1+\eps_2) \left(\frac{231 \,q^3}{256 \,a^{10}}
+\frac{7371 \,q^4}{512 \,a^{14}} + \dots\right)
\\  & +
\eps_1 \eps_2 (\eps_1+\eps_2)^3 \left(-\frac{735 \,q^3}{64\,a^{10}}
-\frac{14469 \,q^4}{128 \,a^{14}} + \dots\right) + \dots
\end{align}

%%%%%%%%%%%%%%%%%%%%%%%%%%%%%%%%%%%%%%%%%%%%%%%%%%%%%%%%%%%%%%%%%%%%%%%%%%%%

\begin{align}
\T{8} =\, &  
2 \,a^8+28 \,a^4 q+\frac{161 \,q^2}{4}+\frac{35 \,q^3}{2 \,a^4}
+\frac{15337 \,q^4}{1024 \,a^8} + \dots
\notag \\ \notag & +
\eps_1 \eps_2 \left(-70 \,a^2 q-\frac{217 \,q^2}{4 \,a^2}-\frac{567 \,q^3}{16 \,a^6}
-\frac{31269 \,q^4}{512 \,a^{10}} + \dots\right)
\\ \notag & +
(\eps_1+\eps_2)^2 \left(105 \,a^2 q+\frac{497 \,q^2}{8 \,a^2}
+\frac{1505\,q^3}{32 \,a^6}+\frac{99561 \,q^4}{1024 \,a^{10}} + \dots\right)
\\ \notag & +
(\eps_1 \eps_2)^2 \left(28 q+\frac{651 \,q^2}{64 \,a^4}
+\frac{1197 \,q^3}{32 \,a^8}+\frac{1318933 \,q^4}{8192 \,a^{12}} + \dots\right)
\\ \notag & +
\eps_1 \eps_2 (\eps_1+\eps_2)^2 \left(-\frac{147 q}{2}
-\frac{385 \,q^2}{16\,a^4}-\frac{4053 \,q^3}{32 \,a^8}
-\frac{1244957 \,q^4}{2048 \,a^{12}} + \dots\right) 
\\ \notag & +
(\eps_1 \eps_2)^3 \left(-\frac{q}{\,a^2}-\frac{649 \,q^2}{128 \,a^6}
-\frac{12021 \,q^3}{256 \,a^{10}}-\frac{6830463 \,q^4}{16384 \,a^{14}} + \dots\right) 
\\  & +
(\eps_1 \eps_2)^2 (\eps_1+\eps_2)^2 \left(\frac{q}{\,a^2}
+\frac{5141 \,q^2}{256 \,a^6}+\frac{137995 \,q^3}{512 \,a^{10}}
+\frac{87543287 \,q^4}{32768 \,a^{14}} + \dots\right) + \dots
\end{align}

\section{Proof of the trace relations in the NS limit}
\la{app:pogho}

We adopt the proposal discussed in \cite{Poghossian:2010pn} for the deformed SW curve in the 
NS limit.
\footnote{We alert the reader that we slightly change the notation compared with \cite{Poghossian:2010pn}.} 
To this aim, we introduce the function $y(z)$ obeying the difference
equation
\be
\la{D.1}
y(z)\,y(z+\hbar)-P(z)\,y(z)+q=0,\qquad P(z) = z^{2}-e^{2}.
\ee
We  solve (\ref{D.1}) perturbatively in $\hbar$ setting 
\be
\la{D.2}
y(z) = \frac{1}{2}\bigg[Y(z)+z^{2}-e^{2}\bigg],\qquad Y(z) = \sum_{n=0}^{\infty}
Y_{n}(z)\,\hbar^{n}.
\ee
The 0th order is the undeformed standard curve (in quartic form with no linear term)
$
Y_{0}(z) = Q(z) \equiv \sqrt{P^{2}(z)-4\,q}.
$
The {deformed SW differential} is then written in terms of 
$
\Psi(z) = \log\left[P(z)+Y(z)\right],
$
and the chiral traces can be extracted from the {resolvent} ${\Psi'(z+\hbar)}$
taking into account the $\hbar$ shift proposed
in  \cite{Poghossian:2010pn}. Plugging into (\ref{D.1}) the expression (\ref{D.2}) with the Ansatz
\be
\la{D.3}
Y(z) = z^{2}+\xi_{1}\,z+\xi_{0}+\xi_{-1}\,\frac{1}{z}+\dots, 
\ee
we determine very easily the coefficients $\{\xi_{n}\}$ and, using  $\bm{u} = 2\,e^{2}$, we find 
\begin{align}
\Psi'(z+\hbar) &= \frac{2}{z}+\frac{\bm{u}}{z^{3}}+\frac{1}{z^{5}}\bigg(
\frac{\bm{u}^{2}}{2}+4\,q
\bigg)+\frac{10\,q\,\hbar}{z^{6}}
+\frac{1}{z^{7}}\bigg(
\frac{\bm{u}^{3}}{4}+6\,q\,\bm{u}+18\,q\,\hbar^{2}
\bigg)\notag\\
&+\frac{1}{z^{8}}\bigg(
21\,q\,\bm{u}\,\hbar+28\,q\,\hbar^{3}
\bigg)+\frac{1}{z^{9}}\bigg(
\frac{\bm{u}^{4}}{8}+6\,q\,\bm{u}+12\,q^{2}+52\,q\,\bm{u}\,\hbar^{2}+40\,q\,\hbar^{4}
\bigg)+\dots.
\end{align}
in agreement with (\ref{3.5}). Notice that the terms in (\ref{3.5}) have the 
same weight under the natural assignment
\be
[\varphi]=1,\qquad [\bm{u}]=2,\qquad [q]=4,\qquad [\hbar]=1.
\ee
Just to give an example, this procedure
gives immediately the next two relations extending the list  (\ref{3.5}). They read
\begin{align}
\T{9} &= 9\,q\,(10\,q+3\,\bm{u}^{2})\,\hbar+108\,q\,\bm{u}\,\hbar^{3}+54\,q\,\hbar^{5}, \notag \\
\T{10} &= \frac{\bm{u}^{5}}{16}+5\,q\,\bm{u}^{3}+30\,q^{2}\,\bm{u}+5\,q\,(86\,q+17\,\bm{u}^{2})\,\hbar^{2}
+200\,q\,\bm{u}\,\hbar^{4}+70\,q\,\hbar^{6},
\end{align}
and turns out to be perfectly satisfied by  the explicit localization result for $\T{9}$ and $\T{10}$.

\section{A cross check by saddle point methods}
\la{app:fucito2}

Trace relations in the NS limit of the  $SU(2)$ $\mc N=2^{\star}$ case
may be treated by the  quantized curve proposed in  \cite{Fucito:2011pn}. 
It allows to deal with the NS limit of the Nekrasov integrals by saddle point methods
working directly at $\eps_{2}=0$.
The practical {perturbative algorithm} is straightforward.
%We begin by defining
%the classical polynomial
%\be
%\la{5.1.8}
%P(x) = x^{2}-a^{2}.
%\ee
%Then, 
We fix an integer $L\ge 1$ and introduce the quantities (where $a_{1} = a = -a_{2}$)
\be
\la{E.1}
x_{u,\ell} = x_{u,\ell}^{(0)}+\sum_{k=\ell}^{L}\xi_{u,\ell}^{(k)}\,q^{k},
\qquad 
x_{u,\ell}^{(0)} = a_{u}+(\ell-1)\,\hbar.
\ee
The set $\{\xi_{u,\ell}^{(k)}\}$ is determined by replacing (\ref{E.1}) into the saddle point equation
\be
\la{E.2}
\left. 1-q\,\frac{w_{L}(x)\,w_{L}(x+\hbar)}{w_{L}(x+m+\hbar)\,w_{L}(x-m)}\right|_{x=x_{u,\ell}}=0.
\ee
where
\be
\la{E.3}
w_{L}(x) = \frac{1}{P(x-L\,\eps)}\,\prod_{u=1}^{2}\prod_{\ell=1}^{L}\frac{x-x_{u,\ell}
-\hbar}{x-x_{u,\ell}}, \qquad P(x) = x^{2}-a^{2}.
\ee
Eq. (\ref{E.2}) must be used as follows: first we evaluate the combination of $w_{L}$ functions with 
generic $x$. Several cancellations occur and after algebraic simplification it is possible to set $x$ to 
each of the $x_{u,\ell}$.
Given $L$, we work out (\ref{E.2}) at order $q^{L}$ included. Increasing $L$, the previous solutions
involving $\xi^{(k)}$ with $k<L$ are unchanged. So the method is iterative, we can start from $L=1$, 
solve the constants $\xi^{(1)}$, use them in the Ansatz with $L=2$ and so on.
After these steps, the chiral traces are simply given by the universal formula
\be
\la{E.4}
\T{n} = \sum_{u}a_{u}^{n}+\sum_{u,\ell}\bigg[
x_{u,\ell}^n-(x_{u,\ell}+\hbar)^{n}-(x^{(0)}_{u,\ell})^{n}+(x_{u,\ell}^{(0)}+\hbar)^{n}
\bigg].
\ee
after the shift $m\to m-\frac{\eps}{2}$. We have checked that (\ref{E.4}) agrees with the previous 
computations up to at least 3 instantons and for $n$ up to 10.  For illustration, let us explain the procedure
 at the simple 1-instanton 
level. The 1-instanton solution (\ref{E.1}) is simply
\be
\la{E.5}
x_{1,1} = a+\frac{\mc C_\hbar (8 a (2 a+\hbar )-\mc C_\hbar)}
{32\, a\, \hbar  (2 a+\hbar )}\,q+\mc O(q^{2}), \ \ 
x_{2,1} = -a-\frac{\mc C_\hbar (8 a (\hbar -2 a)+\mc C_\hbar)}
{32 a \hbar  (2 a-\hbar )}\,q+\mc O(q^{2}).
\ee
Writing $a$ in terms of $\bm{u}\equiv \T{2}$ and applying (\ref{E.4}) we recover the same results as 
from Nekrasov calculation, {\em i.e.} {for even $n$}, we find
\begin{align}
\la{E.6}
\vev{\mbox{Tr}\, e^{z\,\varphi}} &= 2\,\cosh\left(\sqrt{\frac{\bm{u}}{2}}\,z\right)-q\,
\frac{\Chbar\,z}{16\,\sqrt{\bm{u}}\,\hbar\,(2\bm{u}-\hbar^{2})}
\bigg[\notag \\
&
\sqrt{2}\, \hbar\,  \sinh \left(\sqrt{\frac{\bm{u}}{2}}\,z\right) \left(\Chbar\,
   \cosh (z \hbar )+\Chbar+8 \left(\hbar^2-2 \bm{u}\right)\right)\notag \\
   &-2 \sqrt{\bm{u}}
   \left(\Chbar-8 \bm{u}+4 \hbar ^2\right) \cosh \left(\sqrt{\frac{\bm{u}}{2}}\,   z\right) \sinh (z \hbar )
\bigg]+\mc O(q^{2}),
\end{align}
with a similar expression for odd $n$. As a check, we can compute the $z\to 0$ expansion of (\ref{E.6})
and we find (dots denote $\mc O(q^{2})$ contributions)
\begin{align}
\la{E.7}
\vev{\mbox{Tr}\, e^{z\,\varphi}} &= 2+\frac{z^{2}}{2!}\,\bm{u}+\frac{z^{4}}{4!}\,
\bigg[
\frac{\bm{u}^2}{2}+\bm{u} \left(-2 \Chbar q+\dots\right)+\frac{1}{4} \Chbar\, q
   \left(\Chbar-8 \hbar^2\right)+\dots
\bigg]\notag \\
&+\frac{z^{6}}{6!}\,\bigg[
\frac{\bm{u}^3}{4}+\bm{u}^2 \left(-3 \Chbar q+\dots\right)+\bm{u} \left(\frac{3}{8}
   \Chbar \left(\Chbar-40 \hbar^2\right) q+\dots\right)\notag \\
   &+q
   \left(\frac{9 \Chbar^2 \hbar^2}{8}-3 \Chbar \hbar^4\right)+\dots
   \bigg]+\mc O(z^{8}),
\end{align}
in agreement with the 1-instanton contribution in (\ref{4.6}).

% ------------------------------------------------------------------------------------------------------------------------

% U(1)

\section{The special case of $U(1)$ theories}
\la{app:u1}

In this Appendix we briefly discuss the special $U(1)$ case. 
Formally, we shall be using the same localization 
expressions valid for $U(N)$.
This is a useful check given the explicit results of \cite{Fujii:2007qe} 
to be discussed in a moment. 

\subsection{The pure gauge theory}

Let us recall the definitions
\be
\la{F.1}
\text{p} = \eps_{1}\,\eps_{2},\qquad
\text{s} = \eps_{1}+\eps_{2}.
\ee
An explicit calculation gives the following {exact} expressions \footnote{
Notice that to compare with \cite{Fujii:2007qe} we need to send $q\to -q$. This sign flip for $U(N)$
theories with odd $N$ will be further discussed later.}
\begin{align}
\la{F.2}
\T{2} &= -2\, q, \\
\T{3} &= -3 \,q \,\s\,, \notag \\
\T{4} &= 2 \,q \left(\,\p\,-2 \,\s^2\right)+6 \,q^2, \notag \\
\T{5} &= 5 \,q \,\s\, \left(\,\p\,-\,\s^2\right)+25 \,q^2 \,\s\,, \notag \\
\T{6} &= \,q \left(-2 \,\p\,^2+9 \,\p\, \,\s^2-6 \,\s^4\right)+\,q^2 
\left(75 \,\s^2-30 \,\p\,\right)-20   \,q^3, \notag \\
\T{7} &= -7 \,q \left(\,\s\, \left(\,\p\,-\,\s^2\right)^2\right)+
\,q^2 \left(196 \,\s^3-182 \,\p\, \,\s\,\right)-154 \,q^3  \,\s\,, \notag \\
\T{8} &= 2 \,q \left(\,\p\,^3-8 \,\p\,^2 \,\s^2+10 \,\p\, \,\s^4-4 \,\s^6\right)\notag \\
&+14 \,q^2 \left(9 \,\p\,^2-52 \,\p\, \,\s^2+34
   \,\s^4\right)+56 \,q^3 \left(5 \,\p\,-14 \,\s^2\right)+70 \,q^4.
\end{align}
Due to the fact that these expressions are polynomials in $q$, it is possible to find simple relations
fully discussed in \cite{Fujii:2007qe}. The authors of \cite{Fujii:2007qe} considered the further special limit
$\eps_{1}=-\eps_{2}=\hbar$, {\em i.e.} $\p=-\hbar^{2}$ and $\s=0$. Then, one has simple relations like 
\be
\T{4}-\hbar^{2}\,\T{2} = 6\,q^{2}, \qquad
\T{6}-5\,\hbar^{2}\,\T{4}+4\,\hbar^{4}\,\T{2} = -20\,q^{3}, \qquad
\ee
and so on.

\subsection{The $\mc N=2^{\star}$ theory}

The $U(1)$ $\mc N=2^{\star}$ has not been considered in \cite{Fujii:2007qe}, but it 
is a simple extension.
Now, the chiral observables $\T{n}$ are not polynomials in $q$. Nevertheless, they are polynomials
in $m$ and $\bm{\eps}$. In particular, we recall the definition
\be
\mc C = 4\,m^{2}-(\eps_{1}-\eps_{2})^{2},
\ee
and find 
\begin{align}
\T{2} &= \frac{\mc C}{48}\,(\E_{2}-1), \qquad 
\T{3} = -\frac{3}{4}\,\mc C\,\s\,\E_{3}, \notag \\
\T{4} &= \frac{\mc C\,(7\,\mc C+16\,(\s^{2}-\p))}{3840}
-\frac{\mc C^{2}}{384}\,\E_{2}+\frac{\mc C\,(\mc C+2\,\p)}{1152}\,\E_{2}^{2}
-\frac{\mc C\,(\mc C-28\,\p+48\,\s^{2})}{11520}\,\E_{4}.\notag
\end{align}
Due to the fact that $\T{2}$ is known in closed form, the $U(1)$ gauge theory is rather trivial 
from the perspective of our investigation.

% ------------------------------------------------------------------------------------------------------------------------

\section{A technical remark}
\la{app:nontrivial}

Let us discuss the non-triviality of (\ref{7.20}) by discussing 
the difficulties that arise in  a brute force attempt to identify the terms in (\ref{7.19})
 from an explicit low instanton calculation.
 Let us write the 1-instanton expression of $\T{2}$ and $\T{4}$ introducing as before the shortcut
${\mc C=4m^{2}-(\eps_{1}-\eps_{2})^{2}}$. We find 
\begin{align}
\la{G.1}
\T{2} &= 2\,a^{2}+\frac{\mc C\,(-16\,a^{2}+\mc C-4\,\p+4\,\s^{2})}{4\,(4\,a^{2}-\s^{2})}\,q+\mc O(q^{2}), \notag \\
\T{4} &= 2\,a^{4}\notag \\
&+\frac{\mc C\,(-96\,a^{4}+6\,a^{2}\,\mc C-8\,a^{2}\,\p-8\,a^{2}\,\s^{2}-\mc C\,\p-\mc C\,\s^{2}
+4\,\p^{2}+8\,\s^{4})}{4\,(4\,a^{2}-\s^{2})}\,q+\mc O(q^{2}).
\end{align}
We can replace $a$ by a series in $q$ enforcing $\T{2}=\bm{u}$. This gives
\be
\la{G.2}
a = \sqrt\frac{u}{2}-\frac{\mc C\,(\mc C-4\,\p+4\,\s^{2}-8\,\bm{u})}{8\,\sqrt{2\,\bm{u}}
\,(2\,\bm{u}-\s^{2})}\,q
+\mc O(q^{2}).
\ee
Replacing this in the expansion of $\T{4}$ we obtain 
\be
\la{G.3}
\T{4} =\frac{{\bm u}^{2}}{2}+\frac{\mc C\,(\mc C\,(\p+\s^{2}-2\,{\bm u})
-4\,(\p^{2}+2\,\s^{4}-2\,\s^{2}\,{\bm u}-4\,{\bm u}^{2}))}
{4\,(\s^{2}-2\,{\bm u})}\,q+\mc O(q^{2}).
\ee
This expression is not polynomial in $\bm{u}$. Nevertheless, with insight, 
we can look for constants $k_{i}$ such that 
\be
\la{G.4}
\T{4}-\bigg(k_{1} \bm{u'}+k_{2}\,{\bm u}^{2}\bigg) = \text{linear in $\bm{u}$}.
\ee
The idea is that the remaining linear r.h.s. will be taken into account by subleading 
contributions to the trace relation. The condition (\ref{G.4}) gives immediately $k_{2}=\frac{1}{2}$.
Looking at the pole at $\bm{u} = \frac{\s^{2}}{2}$ in the one-instanton contribution, we fix 
$k_{1} = -\p$, in agreement with (\ref{7.19}). If we now move to $\T{6}$, the one-instanton
expression is 
\begin{align}
\la{G.5}
\T{6} &= \frac{1}{4}\,{\bm u}^{3}+\frac{\mc C}{8\,(\s^{2}-2\,{\bm u})}\bigg[
48 {\bm u}^3-6 {\bm u}^2 \left({\mc C}+16 \p-36 \s^2\right)\notag \\
&+{\bm u} 
\left(15 {\mc C} \p-15 {\mc C} \s^2-44 \p^2+48
   \p \s^2-72 \s^4\right)\notag \\
   &+\left(-2 {\mc C} \p^2-6 {\mc C} \p \s^2+9 {\mc C} \s^4+8 \p^3+16 \p^2
   \s^2-24 \s^6\right)
\bigg]\,q+\mc O(q^{2})
\end{align}
Again, we can look for special simplifications in 
\be
\la{G.6}
\T{6}-\bigg(k_{1} \bm{u''}+k_{2}\,\bm{u\,u'}+k_{3} {\bm u}^{3}\bigg) 
\ee
The choice $k_{3}=\frac{1}{4}$ cancels the cubic term $\sim \bm u^{3}$. Vanishing of the residue in the 
one-instanton term gives only the constraint
\be
\la{G.7}
k_{2} = \frac{-4\,k_{1}+4\,\p^{2}-3\,\p\,\s^{2}}{2\,\s^{2}}.
\ee
Thus, {for $\T{6}$ we need the two instanton expression.} After some work, we see that in order to 
cancel the most singular term in the two instanton contribution around the 
pole at $\bm{u}=\frac{\s^{2}}{2}$
we need $k_{1}=-\p^{2}$ and therefore, from (\ref{G.7}), we get $k_{2}= -\frac{3}{2}\,\p$.
All this is in agreement with (\ref{7.19}). However, the meaning of this exercise 
is to show that 
the leading terms captured by (\ref{7.20}) cannot be obtained for generic $n$ by 
means of a fixed instanton
calculation. The order of the expansion must increase as $n$ grows.

\bibliography{N2-Biblio}
\bibliographystyle{JHEP}

\end{document}